\documentclass{iopart}
\usepackage{iopams,setstack}
\usepackage{color}
\usepackage{graphicx}
\usepackage[colorlinks=true]{hyperref}
\newcommand{\LQCD}{\Lambda_{\rm QCD}}
\newcommand{\Nc}{N_{\rm c}}

\newcommand{\Qs}{Q_{\rm s}}
\newcommand{\mux}{\mu_{\rm x}}
\newcommand{\alphas}{\alpha_{\rm s}}
\newcommand{\PT}{P_{\rm T}}
\renewcommand{\PL}{P_{\rm L}}
\newcommand{\mD}{m_{\rm D}}
\newcommand{\rmx}{\mathrm{x}}

\newcommand{\GeV}{\,\mbox{GeV}}
\newcommand{\TeV}{\,\mbox{TeV}}
\newcommand{\fm}{\,\mbox{fm}}

\newcommand{\x}{\mathrm{x}}
\newcommand{\bx}{\boldsymbol{x}}
\newcommand{\bxt}{\boldsymbol{x}_\perp}
\newcommand{\by}{\boldsymbol{y}}

\newcommand{\bp}{\boldsymbol{p}}

\newcommand{\bq}{\boldsymbol{q}}
\newcommand{\bk}{\boldsymbol{k}}
\newcommand{\bkt}{\boldsymbol{k}_\perp}

\newcommand{\calA}{\mathcal{A}}
\newcommand{\calB}{\mathcal{B}}
\newcommand{\calD}{\mathcal{D}}
\newcommand{\calE}{\mathcal{E}}
\newcommand{\calF}{\mathcal{F}}
\newcommand{\calM}{\mathcal{M}}
\newcommand{\calO}{\mathcal{O}}
\newcommand{\calP}{\mathcal{P}}
\newcommand{\calR}{\mathcal{R}}
\newcommand{\dcalA}{\delta\mathcal{A}}
\newcommand{\calL}{\mathcal{L}}

\newcommand{\Wx}{W_{\x}}
\newcommand{\half}{{\textstyle \frac{1}{2}}}

\newcommand{\one}{1\kern-0.36em1}
\begin{document}

%%%%%%%%%%   Title Page   %%%%%%%%%%

%% preprint number %%
\title[Evolution to the Quark-Gluon Plasma]
      {Evolution to the Quark-Gluon Plasma}

\author{Kenji Fukushima}

\address{Department of Physics,
         The University of Tokyo, 7-3-1 Hongo,
         Bunkyo-ku, Tokyo 113-0033, Japan}
\ead{fuku@nt.phys.s.u-tokyo.ac.jp}

\begin{abstract}
  Theoretical studies on the early-time dynamics in the
  ultra-relativistic heavy-ion collisions are reviewed including
  pedagogical introductions on the initial condition with small-$\rmx$
  gluons treated as a color glass condensate, the bottom-up
  thermalization scenario, plasma/glasma instabilities, basics of some
  formulations such as the kinetic equations and the classical
  statistical simulation.  More detailed discussions follow to make an
  overview of recent developments on the fast isotropization, the
  onset of hydrodynamics, and the transient behavior of momentum
  spectral cascades.
\end{abstract}

\submitto{\RPP}
\maketitle

\tableofcontents
\markboth{Evolution to the Quark-Gluon Plasma}%
{Evolution to the Quark-Gluon Plasma}

%%%%%%%%%%   Introduction   %%%%%%%%%%
\section{Introduction}
Early thermalization is the last and greatest unsolved problem in the
ultra-relativistic heavy-ion collisions that have aimed to create a new
state of matter out of quarks and gluons, i.e.\ a state called
Quark-Gluon Plasma (QGP).  As a consequence of non-perturbative and
non-linear nature of the ``strong interaction'', quarks and gluons and
any colored excitations in general cannot be detected directly in
laboratory experiments, which is an intuitive description of the color
confinement phenomenon:  quarks and gluons must be confined into
color-singlet hadrons such as mesons and baryons.  If the temperature
$T$ is comparable to the typical scale of the strong interaction, i.e.\
$\LQCD\sim 0.2\GeV$ ($\sim 2\times 10^{12}\,\mbox{K}$), however,
fundamental degrees of freedom should become more relevant and we may
be able to probe some properties of hot and dense matter with quarks
and gluons manifested.  Then, such ambitious dreams to create a QGP
by our hands have motivated the installation of high-energetic beams
(see an essay \cite{Baym:2001in} about two decades
\textit{from dreams to beams}).   In fact, an extraordinarily
high-energetic collision of two nuclei is a unique tool to realize
such high energy density and temperature.  It is widely believed that
our wish to create the QGP has been successfully granted at
Relativistic Heavy-Ion Collider (RHIC) and more activities at even
higher energies are continued to Large Hadron Collider (LHC).  There
are, however, still some disputes about physical characteristics of
the QGP from the theoretical point of view.  All subtleties come from
lack of clear-cut definition of the QGP from the first-principle
theory of the strong interaction, i.e.\ quantum chromodynamics (QCD).

Perturbative calculations based on QCD have been established as
theoretical descriptions in terms of quasi-particles of quarks and
gluons (or ``partons'' collectively).  Although there is no order
parameter for a change from the hadronic phase to the partonic phase,
we may well give a working definition of the QGP as a state that
satisfies following (at least) two conditions.  First, the physical
degrees of freedom should be partons rather than hadrons, so that
perturbative QCD (pQCD) can be a good description of the system.
Second, the created state should form \textit{matter} unlike a simple
superposition of each partonic reaction.  For this latter condition,
for decades conventionally, a far stronger condition of
\textit{thermalization} had been imposed.  Precisely speaking, local
thermal equilibrium (LTE) had been assumed to link theoretical
modeling to experimental QGP signatures.  It is, however, very hard to
account for the LTE with QCD microscopic processes within a time scale
$\lesssim\LQCD^{-1}\sim 1\fm/c$.  Eventually, after many trials and
errors (one of earliest discussions can be found in \cite{Baym:1984np}
and the difficulty was revisited in \cite{Kovchegov:2005ss}),
theoretical ideas went around came around to the very starting point
-- what is \textit{matter} at all?  This issue is sometimes discussed
in the context of the origin of \textit{collectivity} of smaller
systems involving proton, deutron, and light ions at LHC energies.

In this review, we do not discuss experimental and phenomenological
studies of collectivity in small systems, which are currently ongoing,
and we still need wait to see an ordered consensus out from disordered
arguments.  Here, we would look over purely theoretical approaches to
reveal real-time QCD dynamics during the evolution to the
QGP.\ \ Fortunately, we can specify the trustworthy initial condition
for the system right after the heavy-ion collisions using our pQCD
knowledge.  It is known that the gluon distribution function has
increasing behavior with increasing reaction energy and classical
color fields give a better description of such an overpopulated state
than individual gluons, which can be understood in analogy to
Weizs\"{a}cker-Williams fields in quantum electrodynamics (QED).  The
theoretical framework with coherent classical color fields (sometimes
called non-Abelian Weizs\"{a}cker-Williams fields
\cite{Kovchegov:1996ty}) is known as the color glass condensate
(CGC).  Thus, we can say that, for a full understanding of the QGP
physics, the missing link is a bridge between the CGC initial
condition and the QGP described well by hydrodynamic equations.  In
other words, using a more general term, we can define our theoretical
question as follows:  How can a full quantum system get to a LTE state
as a solution of the initial value problem starting with coherent
fields?

Limiting our considerations to a specific situation in the
relativistic heavy-ion collision, we can categorize the issues of
thermalization into three distinct (and probably related)
characterizations --- isotropization, hydrodynamization, and spectral
cascades.  Let us briefly address them in order.  The first is the
(partial) \textit{isotropization}.  In the case of the heavy-ion
collision, the system is expanding in time and the interaction should
be turned off for a dilute system as long as we can neglect running
effects of the strong coupling constant or confining forces.  Such a
theoretically idealized limit of non-interacting quarks and gluons in
an expanding box is often called the free-streaming limit.  The
isotropization problem is an issue of how to explain the fact that the
system can resist against a tendency falling into the free-streaming
limit especially when the system is expanding.  The second is the
onset where hydrodynamic equations start working well to capture the
real-time evolution of the system.  In some literature this onset is
discussed under the name of the hydronization or
\textit{hydrodynamization}.  If the system sits in the LTE state, the
hydrodynamic model should be valid, and in this sense, the LTE is a
sufficient condition but not a necessary one for hydrodynamization.
Therefore, we may take the switching time to hydrodynamics earlier
than the genuine LTE time.  Recent developments include a significant
extension of the hydrodynamic regime once higher-order derivative
(dissipative) terms are implemented.  If we knew some optimal
resummation scheme, the hydrodynamic equations may have a validity
region even in the vicinity of the coherent initial conditions.  The
third is a dynamical evolution toward the thermal spectrum in momentum
space.  A very classical problem along this line is found in the
asymptotic solution of the quantum Boltzmann equation.  The detailed
balance is satisfied with the Bose-Einstein distribution for bosons
and the Dirac-Fermi distribution for fermions.  Once those thermal
spectra appear, the physical temperature is well defined, and the LTE
is fully justified.  This kind of analysis can provide us with
thorough information on the thermalization problem, namely, the whole
temporal profile of the distribution functions (possibly with some
forms of condensates).  More interestingly, besides, a non-trivial and
intriguing question is whether any type of stable solution other than
thermal spectra can be possible or not.  Thermal distribution
functions show exponential damping at large momenta and the
temperature is nothing but a slope parameter to characterize how fast
this exponential decrease is.  In some physical circumstances like a
turbulent flow, before reaching such an exponential shape, a power-law
type of distribution may appear as a consequence of
\textit{spectral cascade} in momentum space.  To reiterate this third
step, our theoretical mission is to seek for a possibility of various
pre-thermalization stages \cite{Berges:2004ce}.

%---   figure   ---%
\begin{figure}
 \begin{center}
 \includegraphics[width=0.8\textwidth]{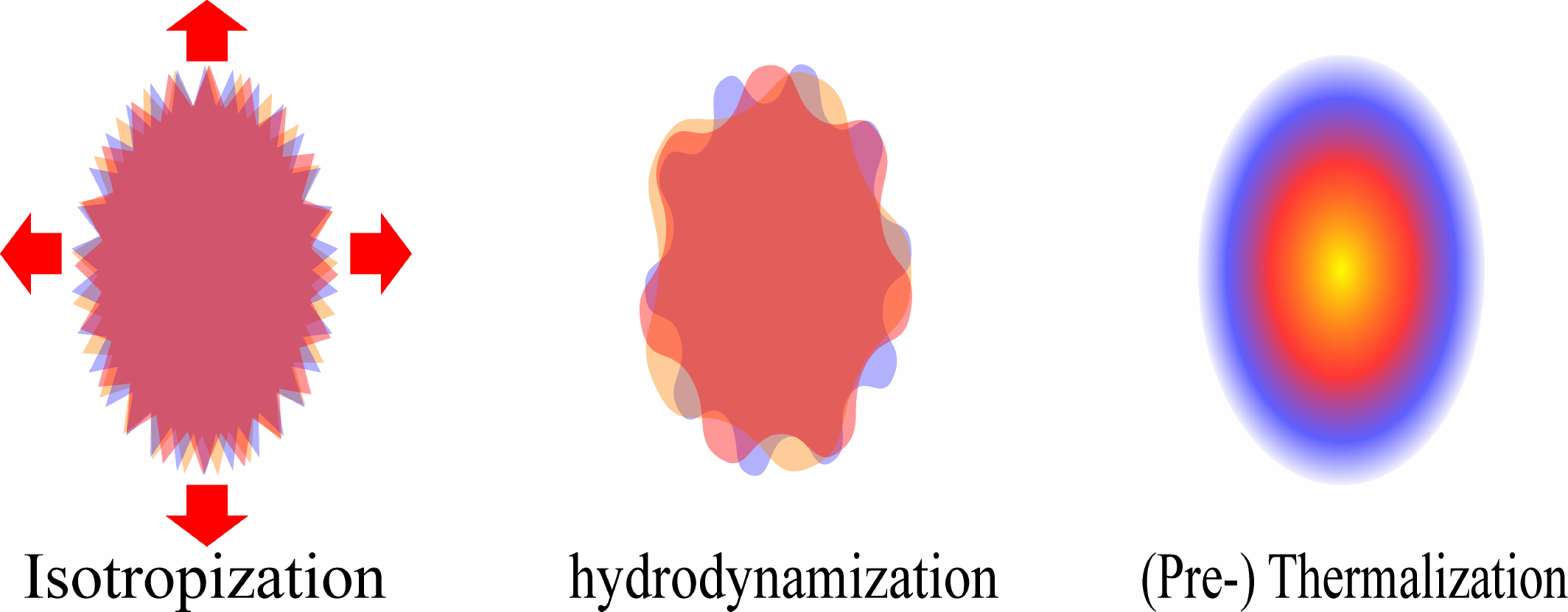}
 \end{center}
 \caption{Schematic (conceptual) illustration of the isotropization
   (left), the hydrodynamization (middle), and the
   (pre-)thermalization (right), which respectively picturize initial
   quantum fluctuations with anisotropic expansion, smoothened
   distributions after some time, and emergence of (pre-)thermal
   spectra.}
 \label{fig:schematic}
\end{figure}
%---   figure   ---%

Figure~\ref{fig:schematic} is a schematic illustration to picturize
ideas of these three steps intuitively.  The left picture of
figure~\ref{fig:schematic} is about the isotropization of the
transverse pressure $\PT$ and the longitudinal pressure $\PL$.  The
ratio $\PL/\PT$ does not have to approach the unity, and nevertheless,
it is expected to converge to a certain value instead of monotonic
decrease to zero.  For the realization of some time window in which
$\PL/\PT$ can be approximately constant, it is crucial to take account
of correct quantum spectrum of initial fluctuations. Although we do
not go into phenomenological challenges in this review, we would note
that $\PL/\PT$ might be (and should be) constrained by a scrupulous
comparison of hydrodynamic simulations and experimental data from the
heavy-ion collisions.  The middle picture of
figure~\ref{fig:schematic} visualizes how the hydrodynamization takes
place.  In principle, hydrodynamic equations are conservation laws,
and they are always useful as long as we are interested in slow
components in real-time dynamics.  For practical purposes, however, we
need close a set of equations to solve them and it should be
reasonable to adopt the hydrodynamic description when spacetime and
momentum variations are sufficiently smoothened (and interactions are
localized).  It is also of pragmatic importance to resolve the
hydrodynamization problem for theorizing hydrodynamics better.  A
recent reformulation named aHydro \cite{Nopoush:2014qba} is a clear
example to extend the validity of hydrodynamics with optimal
resummation.  The right picture of figure~\ref{fig:schematic} sketches
emergence of some scaling solution that could be identified as a
signature of pre-thermalization.  There are several theoretical
speculations such as the turbulent spectrum, the non-thermal fixed
point, and the inverse Kolmogorov cascade and the resultant formation
of a Bose-Einstein condensate (BEC), and so on, together with
numerical demonstrations.  It is, so far, not very obvious how these
scenarios may or may not have an impact on heavy-ion collision
phenomenology.  The most serious problem lies in a technical
difficulty in estimating the relevant time scale.  Almost all
simulations seem to require unphysically long time, hinting that
something important may be still missing.

This review is organized in the following manner.  After this
Introduction, in \sref{sec:foundations}, we will elucidate some
theoretical foundations for readers who would like to learn quickly
what ideas were discussed in the past and what problems still remain
today.  We will start with a pedagogical introduction to the CGC
theory and explain characteristic features of the CGC-type initial
conditions in \sref{sec:cgc}.  Then, as a classic example of
CGC-based arguments of thermalization, in \sref{sec:bottomup}, we will
introduce what is called the ``bottom-up scenario'', which underlies
all thermalization ideas in contemporary approaches.  We also briefly
mention on the plasma and glasma instabilities afterward.  In
\sref{sec:formulations} we discuss several theoretical methods for the
real-time quantum simulation.  In fact, unlike lattice discretized
quantum field theories in Euclidean spacetime for which the
Monte-Carlo sampling is useful, there is no general non-perturbative
algorithm applicable for Minkowskian spacetime.  What we can do with
QCD at best is to take some limit so that a particular approximation
can be validated.  In the dilute limit especially when a
quasi-particle approximation makes sense, the kinetic equation is the
most powerful tool even for QCD and, in principle, systematic
estimation of the collision term is perturbatively doable, though
the numerical calculation becomes desperately heavier with higher
order terms.  We will flash an earliest argument of thermalization by
means of the Boltzmann equation in the relaxation time approximation.
In the opposite case of the dense and overpopulated limit, the
semi-classical approximation would be a natural choice of most suited
descriptions, which consists of the solution of the classical
equations of motion and the Wigner function.  In quantum mechanics the
semi-classical approximation works for many problems, but for quantum
field theories, the semi-classical approximation or the classical
statistical simulation has delicate subtleties affected by ultraviolet
(UV) modes for which the density is small and the approximation
inevitably breaks down.  In \sref{sec:unconventional} we will also
give very short remarks on some unconventional approaches such as the
Kadanoff-Baym equations, stochastic quantization, and gauge/gravity
correspondence.  Successful examples for specific problems with these
techniques exist and there may be some potential for the future, but
so far the applicability is limited to rather academic
considerations.

We will continue to \sref{sec:isotropization},
\sref{sec:hydrodynamics}, and \sref{sec:thermal} to go into more
detailed discussions on the issues of the isotropization, the
hydrodynamization, and the pre-thermalization, respectively.  We put
our emphasis on the self-contained derivations of more or less
established physics in \sref{sec:foundations}, while in later sections
we will pick up and outline some of most recent results.
Specifically, we will mainly focus on selected results on the
classification of scaling solutions, the success of the aHydro
formulation, and the speculative scenario of a gluonic BEC formation.
Readers interested in hydrodynamic simulations together with a comparison to
heavy-ion data can consult a recent review \cite{Hirano:2012kj}.
Because the thermalization problem is a rapidly growing subject, new
progresses are steadily reported.  We will not try to make this review
comprehensive in vain but will take a more pragmatic strategy to
explicate the problems and the progresses rather than to give an
answer.  For the most state-of-the-art outcomes, readers are
encouraged to study further with proceedings contributions for Quark
Matter conference series.

%%%%%%%%%%   Theoretical Foundations   %%%%%%%%%%
\section{Theoretical Foundations}
\label{sec:foundations}
We will exposit some theoretical formulations based on QCD that are
useful to quantify microscopic processes of the evolution to the
QGP.\ \ The early time dynamics in the heavy-ion collision has a
universal scale called the saturation momentum (denoted as $\Qs$)
apart from the typical QCD scale, $\LQCD$.  So, it is indispensable to
implement $\Qs$ properly for modern approaches to the thermalization
problem.

%%%%%   Small-x physics and Color Glass Condensate   %%%%%
\subsection{Small-x Physics and Color Glass Condensate}
\label{sec:cgc}
An old-fashioned quark model tells us that the nucleon is composed
from three valence quarks.  Such a naive picture could hold, however,
for the net quantum number only and there should be a far richer
structure with sea quarks and gluons once quantum corrections are
included.  In the infinite momentum frame in which the nucleon has an
infinitely large momentum, the life time of virtual excitations is
elongated due to Lorentz time dilatation, so that the parton
distribution functions including virtual excitations become
well-defined physical observables.  A parton with a large momentum can
radiate softer partons one after another in quantum processes, and
there should be more abundant partons with smaller momenta.  To
quantify this, it is convenient to introduce Bjorken's x that is a
fraction of the longitudinal momentum carried by a parton over the
total momentum of a projectile.  According to the data from Hadron
Electron Ring Accelerator (HERA) the gluon distribution function is
about twenty times larger than the quark distribution function already
around $\rmx\sim 10^{-2}$, and in the first approximation, we can
neglect contributions from quarks.

For the thermalization problem, we should consider processes involving
soft momenta $\lesssim 1\GeV$ and then the relevant $\rmx$ is roughly
$\rmx\sim 10^{-2}$ for RHIC energy of $200\GeV$/nucleon and
$\rmx\sim 10^{-3}$ for LHC energy of $5.5\TeV$/nucleon.  In this
small-$\rmx$ regime, we can safely limit our considerations to gluonic
contributions only using the pure Yang-Mills theory instead of full
QCD with dynamical quarks.  Further simplification occurs at
sufficiently small $\rmx$:  when the gluon distribution function
$G(\rmx,Q)$ where $Q$ represents the transverse momentum is such
enhanced, gluons eventually saturate the transverse area $\pi R_A^2$
of the nucleon or nucleus.  We should note that this happens in a way
dependent on $\rmx$.  Actually, the transverse size of the probed
parton is characterized by $Q^{-1}$ in the Breit frame and thus the
corresponding interaction cross section is $\sim \alphas\Nc Q^{-2}$.
Then, the saturation condition reads:
$\alphas\Nc G(\rmx,\Qs) \Qs(\rmx)^{-2}/(\Nc^2-1) \simeq \pi R_A^2$.  It
is obvious that the left-hand side is the total cross section per one
color.  The solution of this equality yields a qualitative definition
of the saturation scale $\Qs(\rmx)$.  The most important
implication from the saturation is that physical quantities should
scale with $\Qs(\x)$ in a universal way.  More concretely, as a
consequence of the saturation, the total cross section
$\sigma_{\gamma^\ast p}(\rmx,Q^2)$ of a proton and a virtual photon
(with an electron vertex amputated) should no longer be a function
of $\rmx$ and $Q^2$ independently but is a function of a scaling
variable $\tau\equiv Q^2/\Qs^2(\rmx)$ only.  Experimental data from
HERA with various combinations of $\rmx$ and $Q^2$ exhibit beautiful
scaling behavior called the ``geometric scaling''
\cite{Stasto:2000er} with the following parametrization;
\begin{equation}
 \Qs^2(\rmx) = Q_0^2 (\rmx/\rmx_0)^{-\lambda} \;,
\label{eq:Qs}
\end{equation}
where $Q_0=1\GeV$ is pre-fixed and $\rmx_0=3.04\times 10^{-4}$,
$\lambda=0.288$ have been determined from the data at
$\rmx<10^{-2}$.  This functional form is also suggested by a
solution of the BFKL equation which is a linear quantum evolution
equation with changing $\rmx$.  \Eref{eq:Qs} provides us with a more
quantitative definition of $Q_s(\rmx)$ used for phenomenological
applications such as the prediction of the hadron multiplicity in a
KLN model \cite{Kharzeev:2004if}.

It should be noted that the saturation is a sufficient condition for
the geometric scaling, but may not be a necessary condition.  This
means that the geometric scaling may hold outside of the saturation
regime and this is indeed the case in view of the experimental data:
not only $\tau\lesssim 1$ but larger $\tau\gtrsim 10^2$ also show
the scaling behavior.  This experimental finding is extremely
important for reality of the CGC;  for $\tau\gtrsim 1$ the parton
transverse size $\sim Q^{-2}$ is certainly smaller than necessary
for the saturation $\sim \Qs^{-2}$.  Therefore, the validity region of
the CGC must be wider than naively expected.  This ``extended
geometric scaling'' could be a consequence from quantum evolution
equations with changing $\rmx$ and $Q^2$ that maintain the geometric
scaling even beyond the saturation regime \cite{Iancu:2002tr}.  In
discussions in what follows throughout this review, we shall require
that the kinematic regions involving $Q^2\sim \Qs^2(\rmx)$ dominate
processes of our interested physics.

In the case of the nucleus-nucleus collision, the transverse parton
density is significantly enhanced with the atomic number $A$.  Because
the nuclear thickness scales with $A^{1/3}$, as compared to the proton
case, $\Qs^2(\x)$ should be accompanied by $A^{1/3}$ which is as a
large factor as $\sim 6$ for gold and lead ions.  This is a
tremendously large factor;  the collision energy is $\sim 27$ times
increased from RHIC to LHC and so relevant $\rmx$ becomes $\sim 1/27$
times smaller.  Using \eref{eq:Qs} we can easily make an estimation
and conclude that $\Qs(\x)$ is increased by a factor $\sim 2.6$ only.
Thus, the CGC regime should be activated much earlier for the
heavy-ion collision than for the proton, and in view of the geometric
scaling in $\sigma_{\gamma^\ast p}$ for $\rmx<10^{-2}$, we can be
confident that the CGC be a trustful description of soft gluons with
momenta $\lesssim 1\GeV$ or even higher.

%%%   CGC effective theory   %%%
\subsubsection{CGC effective theory}
The general strategy to obtain an effective theory is to integrate
unwanted degrees of freedom out.  We can consider an effective theory
for soft gluons by regarding $\rmx$ as a separation scale of hard and
soft gluons.  It has been shown that integrating hard gluons out leads
to a \textit{classical} color source $\rho$ for soft gluons.  In such
a way the probability function $\Wx[\rho]$ that characterizes how
$\rho$ is distributed evolves with changing $\rmx$, and the evolution
of $\Wx[\rho]$ should follow from a renormalization group equation.
This is actually a contemporary derivation of the BFKL equation not
from each Feynman diagram but from the invariance of the partition
function \cite{JalilianMarian:1997jx} and its non-linear extension,
i.e., the JIMWLK equation
\cite{Iancu:2000hn,Ferreiro:2001qy,Hatta:2005rn,Fukushima:2006cj} was
derived as an extension of this method.

Soft gluons are thus given by a classical solution of the Yang-Mills
equations of motion sourced by $\rho$ whose distribution is dictated
by $\Wx[\rho]$.  In a frame where the proton or the heavy-ion is moving
at the speed of light in the positive $z$ direction, the color source
is static in terms of the light-cone time, i.e.\ $\rho=\rho(x^-,\bx)$
where $x^\pm=(t\pm z)/\sqrt{2}$ and $\bx$ refers to the 2-dimensional
transverse coordinates.  The Yang-Mills equations to be solved then
read:
\begin{equation}
 \calD_\mu \calF^{\mu\nu} = \delta^{\nu +} \rho(x^-,\bx) \;.
\label{eq:one_source}
\end{equation}
In this review we consistently use calligraphy letters to represent
classical fields.  We here work in the light-cone gauge with
$\calA^+=0$ and we \textit{assume} $\calA^-=0$ to solve
\eref{eq:one_source}.  Then, let us take a static color rotation to
gauge $\calA^i$ away.  Because of $x^+$ independence, such a gauge
rotation $V(x^-,\bx)$ does not affect $\calA^-=0$ (which is confirmed
from $V^\dag \partial^- V=0$, where we should note that
$\partial^-=\partial_+=\partial/\partial x^+$).  In this rotated color
basis, hence, \eref{eq:one_source} is reduced to the standard
2-dimensional Poisson equation for $\calA^+$ and it is easy to find
the solution as $\calA^+(x^-,\bx) = -\bnabla^{-2} \rho(x^-,\bx)$
\cite{Kovchegov:1996ty}.  We can immediately rotate this solution back
to the light-cone gauge using the rotation matrix $V$ and finally we
arrive at the following solution:
\begin{equation}
 \calA^i = \alpha^i \equiv -\frac{1}{\rmi g}V(x^-,\bx)\partial^i
  V^\dag(x^-,\bx)\;, \qquad
 \calA^\pm = 0\;,
\label{eq:Ai}
\end{equation}
where the rotation matrix to eliminate $\calA^+$ is found to be
\begin{equation}
 V^\dag(x^-,\bx) = \calP \exp\biggl[-\rmi g\int_{-\infty}^{x^-}
  \rmd \xi^-\; \bnabla^{-2}\rho(\xi^-,\bx)\biggr]\;.
\end{equation}
Here, $\calP$ stands for the time ordering.  Now we are ready to
compute physical observables such as the energy-momentum tensor given
in terms of $\alpha^i$.  We can write the expectation value of an
arbitrary operator $\calO[\alpha^i]$ (for example,
$\calO[\alpha^i]=\tr[V(\infty,\bx)V^\dag(\infty,\by)]$ for a dipole
scattering amplitude) down as follows:
\begin{equation}
 \langle \calO[\alpha^i]\rangle = \int \rmd\rho\,
  \Wx[\rho]\,\calO[\alpha^i] \;.
\label{eq:average}
\end{equation}
The above-mentioned calculational scheme with classical fields
$\calA^\mu$ and the weight function $\Wx[\rho]$ is commonly referred
to as the color glass condensate or CGC (for a review; see
\cite{Gelis:2012ri});  in the first approximation $\rho$ is a random
\underline{color} source, which is reminiscent of the theory of spin
\underline{glass}, and is described by classical fields as if they were
\underline{condensates} in scalar theories prescribed by the
Gross-Pitaevskii equation, which explains the name of the color glass
condensate.

It should be noted that solving the classical equations of motion is
an efficient resummation technique to take account of infinite Feynman
diagrams at once, especially for a special case when both terms in the
covariant derivative, $\calD_\mu=\partial_\mu-ig\calA_\mu$, are
comparable.  In the CGC regime, actually, $\partial_\mu$ picks up an
energy and momentum scale $\sim \Qs$.  Also, the color source should
be as large as $\rho\sim \Qs/g$ and thus $\calA_\mu\sim \Qs/g$.  Then,
the perturbation theory must be reorganized not around the vacuum but
around the CGC background fields $\calA_\mu$.  Such reorganized
perturbative calculations result in the renormalization group flow of
$\Wx[\rho]$ and the presence of $\calA_\mu$ makes an upgrade of the
BFKL equation into the JIMWLK equation.  We should note that
perturbative calculations can be useful for $\partial_{\x}\Wx[\rho]$
but cannot figure out $\Wx[\rho]$ itself.  So, we need to rely on some
empirical parametrization for $\Wx[\rho]$ at some initial $\rmx$.  The
simplest choice is a Gaussian Ansatz
\cite{McLerran:1993ni,Iancu:2002aq}, that is;
\begin{equation}
 \Wx[\rho] = \exp\biggl[ -\int\rmd^2\bx\,\rmd x^-\,
  \frac{|\rho(x^-,\bx)|^2}{2g^2\mux^2(x^-)} \biggr]\;.
\label{eq:MV}
\end{equation}
In terms of a color component, $\rho=\rho^a t^a$ where $t^a$ is an
element of color-group algebra in the fundamental representation, the
above Gaussian form is equivalent to requiring the two-point function
as $\langle\rho^a(x^-,\bx)\rho^b(y^-,\by)\rangle
=g^2\mux^2(x^-)\,\delta^{ab}\,\delta(x^--y^-)\,\delta^{(2)}(\bx-\by)$.
This choice of the weight function in \eref{eq:MV} defines what is
known as the McLerran-Venugopalan (MV) model and, naturally, a unique
scale $\mux(x^-)$ is related to $\Qs(\rmx)$:  parametrically
$\Qs\sim g^2\mux$ so that $\calA_\mu\sim \Qs/g$.  Typically $\mux$ is
chosen around $1\GeV$ for RHIC and $2$-$3$ times greater for
LHC.\ \ The Gaussian choice has an advantage that we can perform
analytical calculations for the color average in \eref{eq:average},
which in most cases simplifies significantly in the large $\Nc$ limit
(see \cite{Fukushima:2007dy} for useful mathematical formulas).

%%%   Initial condition ...   %%%
\subsubsection{Initial condition for the relativistic heavy-ion collision}
\begin{figure}
 \begin{center}
 \includegraphics[width=0.4\textwidth]{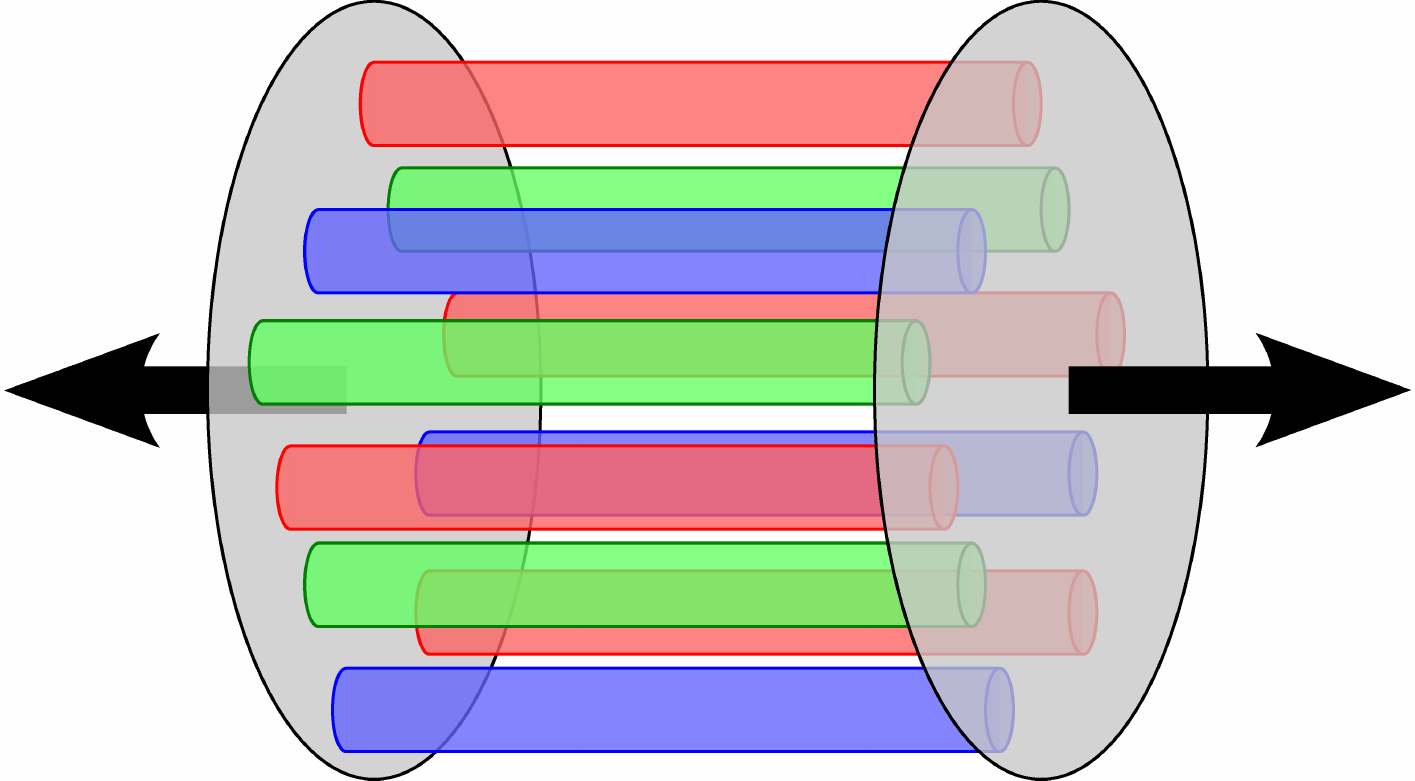}
 \end{center}
 \caption{Schematic illustration of the glasma initial condition for
   the heavy-ion collision.  Longitudinal color electric and magnetic
   fields stretch between two nucleus sheets forming a structure with
   color flux tubes.  Figure is taken from \cite{Fukushima:2011ca}.}
 \label{fig:glasma}
\end{figure}

The same idea of saturation physics can be applied to the relativistic
heavy-ion collision and in this case both the target and the
projectile are dense objects.  To take full account of non-linear
color fields from both nuclei, the Yang-Mills equations that we must
solve read:
\begin{equation}
 \calD_\mu \calF^{\mu\nu} = \delta^{\nu+}\rho^{(1)}(x^-,\bx)
  + \delta^{\nu-}\rho^{(2)}(x^+,\bx)\;.
\label{eq:two_source}
\end{equation}
Here (1) and (2) in the upper subscript refer to the nuclei moving in
the positive and the negative $z$ directions, respectively.  Unlike the
single-source problem in \eref{eq:one_source}, we cannot generally
solve \eref{eq:two_source} in an analytically closed form.  In the
spacelike regions two sources cannot communicate with each other
because of causality, and so the problem is to be reduced to the
one-source problem.  Imposing continuity from these solutions, we can
at best write the analytical solution down on the light cone.  For the
description of the heavy-ion collision, the Bjorken coordinates
$(\tau,\eta)$ are more useful than the light-cone coordinates $x^\pm$,
which are related as
\begin{equation}
 \sqrt{2}x^\pm = \tau\, \rme^{\pm\eta} \;.
\end{equation}
Then, in the radial gauge $\calA_\tau=x^-\calA^++x^+\calA^-=0$, the
solution of \eref{eq:two_source} on the light cone at $\tau=0$ takes a
form of \cite{Kovner:1995ja}
\begin{eqnarray}
 && \calA_i = \alpha^{(1)}_i + \alpha^{(2)}_i \;,\qquad
 \calA_\eta = 0\;,\nonumber\\
 && \calE^i = 0\;,\qquad
 \calE^\eta = \rmi g\Bigl( \bigl[ \alpha_1^{(1)},\alpha_1^{(2)} \bigr]
  + \bigl[ \alpha_2^{(1)},\alpha_2^{(2)} \bigr] \Bigr)\;,
\label{eq:initial}
\end{eqnarray}
where $\calE^i$ and $\calE^\eta$ are the transverse and the
longitudinal components of the classical color electric fields.  It is
quite intuitive that $\calA_i$ is just a linear superposition of
$\alpha_i^{(1)}$ and $\alpha_i^{(2)}$, while $\calE^\eta$ appears from
the non-Abelian character and there is no counterpart in QED.\ \ With
this initial condition \eref{eq:initial}, we should solve the
Yang-Mills Hamilton equations in the Bjorken coordinates:
\begin{equation}
 \partial_\tau\calE^i = \frac{1}{\tau}\calD_\eta \calF_{\eta i}
  + \tau\calD_j\calF_{ji}\;,\quad
 \partial_\tau\calE^\eta = \frac{1}{\tau}\calD_j \calF_{j\eta}\;,
\label{eq:eom}
\end{equation}
with the canonical conjugate momenta defined ordinarily by
\begin{equation}
 \calE^i = \tau\partial_\tau \calA_i\;,\qquad
 \calE^\eta = \frac{1}{\tau}\partial_\tau\calA_\eta\;.
\label{eq:eom2}
\end{equation}
We should note that $\calE^\eta$ has a correct mass dimension of the
electric field but $\calE^i$ does not.  In physical terms
$\calE^i/\tau$ should be interpreted as the genuine transverse
electric field which also goes to zero in the $\tau\to0^+$ limit.
Using $\calA_i$ in \eref{eq:initial} we can readily calculate the
initial color magnetic field as
\begin{equation}
 \calB^i = 0\;,\qquad
 \calB^\eta = \calF_{12} = -\rmi g\Bigl(
  \bigl[\alpha^{(1)}_1,\alpha^{(2)}_2\bigr]
 +\bigl[\alpha^{(2)}_1,\alpha^{(1)}_2\bigr] \Bigr)
\label{eq:initialB}
\end{equation}
using the fact that $\alpha_i^{(n)}$ is a pure gauge and so its field
strength is vanishing.  Although the combinations of indices for
initial $\calE^\eta$ in \eref{eq:initial} and initial $\calB^\eta$ in
\eref{eq:initialB} are slightly different, the squared expectation
values turn out to be identical after taking the color average with
the Gaussian weight as defined in \eref{eq:MV}.  These identical
$\langle \calE^\eta \calE^\eta\rangle$ and
$\langle \calB^\eta \calB^\eta\rangle$ lead us to a very suggestive
profile of the initial condition for the heavy-ion collision as
illustrated in \fref{fig:glasma}.

%---   figure  ---%
%---   figure   ---%

The evolving color fields starting with the initial condition in
\eref{eq:initial} are the foundation of the ``glasma'' (named in
\cite{Lappi:2006fp} though its physics was known traced back to the
Lund string model) which is a transient
state between the color \underline{glass} condensate and the
quark-gluon \underline{plasma} -- glasma as a coined word from them.
The most essential property associated with the glasma initial
condition is, as sketched in \fref{fig:glasma}, the presence of
longitudinal color electric and magnetic fields with boost invariance
(i.e.\ $\eta$ independence), which may be a source for rapidity
correlation (ridge structure) \cite{Dumitru:2008wn} and also local
parity violation \cite{Kharzeev:2001ev}.  Because $\Qs$ is the
universal scale, each color flux tube is expected to be localized in a
domain whose transverse extent is $\sim \Qs^{-1}$.  In the MV model,
however, it is very difficult to see such a structure by eyes.
Recently the correlation length possibly related to the flux tube
structure has been numerically measured in the MV model by means of
spatial Wilson loops and the color flux tube picture has been
partially verified \cite{Dumitru:2013wca}.

For our present consideration on the thermalization problem, it is
critically important to recognize that the longitudinal pressure is
inevitably negative with this type of glasma initial condition.  We
can understand such a negative pressure intuitively:  the longitudinal
fields have positive energy density and so it would cost a more
positive energy to stretch the color flux tubes farther.  This implies
that two nucleus sheets feel an attractive force to decrease the flux
tube energies, leading to a negative pressure.  On the algebraic level
we can see this from
\begin{eqnarray}
 && \PT \equiv \frac{1}{2}\langle T^{xx} + T^{yy}\rangle
  = \Bigl\langle \tr\bigl[ \calE^{\eta a}\calE^{\eta a}
     + \calF_{12}^a \calF_{12}^a \big]\Bigr\rangle \;,
\label{eq:PT}\\
 && \PL \equiv \langle \tau^2 T^{\eta\eta}\rangle
  = \frac{1}{\tau^2} \Bigl\langle \tr\bigl[ \calE^{ia}\calE^{ia}
     + \calF_{\eta i}^a\calF_{\eta i}^a \bigr]\Bigr\rangle
    - \PT \;.
\label{eq:PL}
\end{eqnarray}
In the initial stage the contribution from transverse fields is
negligibly small (regardless of $1/\tau^2$), and so $\PT>0$ and
$\PL\sim -\PT<0$ should be simultaneously developing for finite but
small $\tau$.

%---   figure  ---%
\begin{figure}
 \begin{center}
 \includegraphics[width=0.45\textwidth]{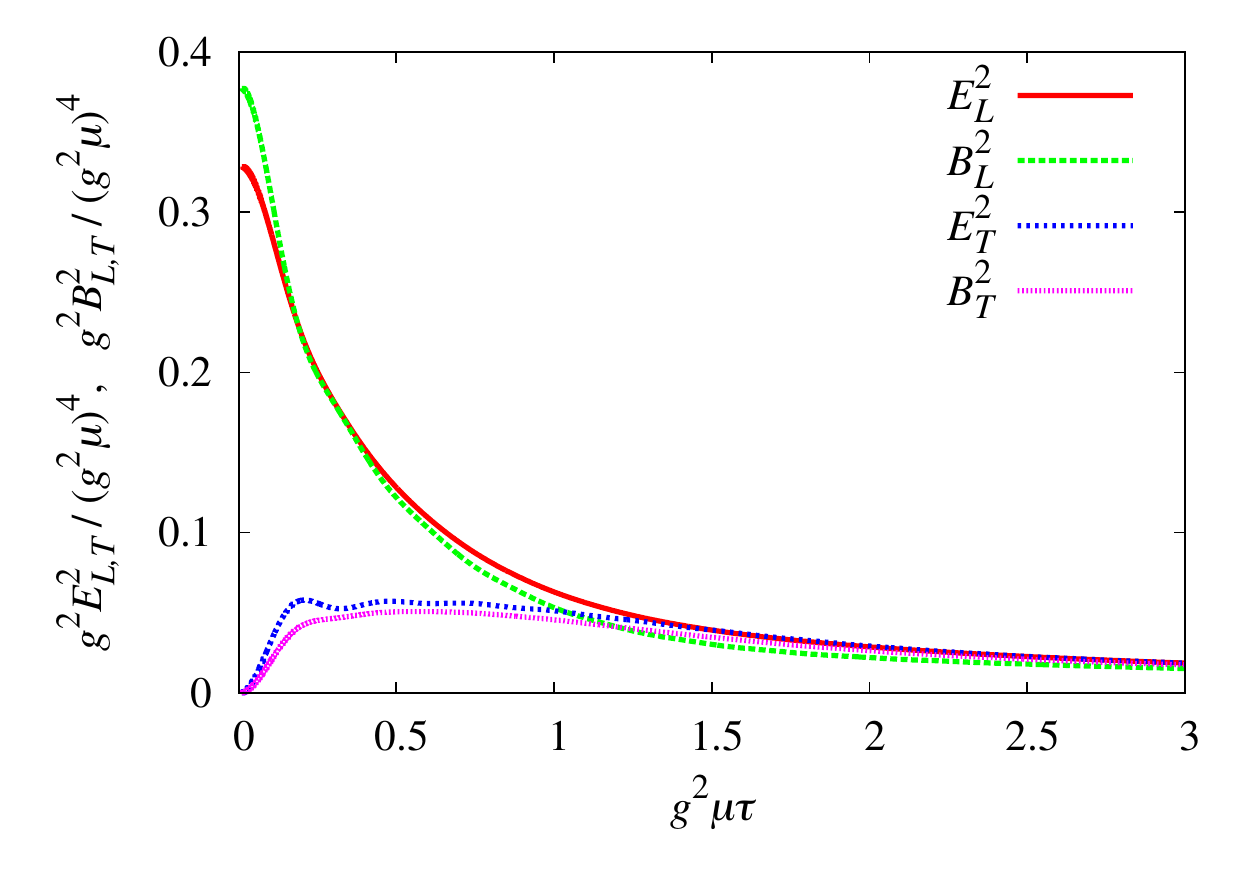}\hspace{1em}
 \includegraphics[width=0.45\textwidth]{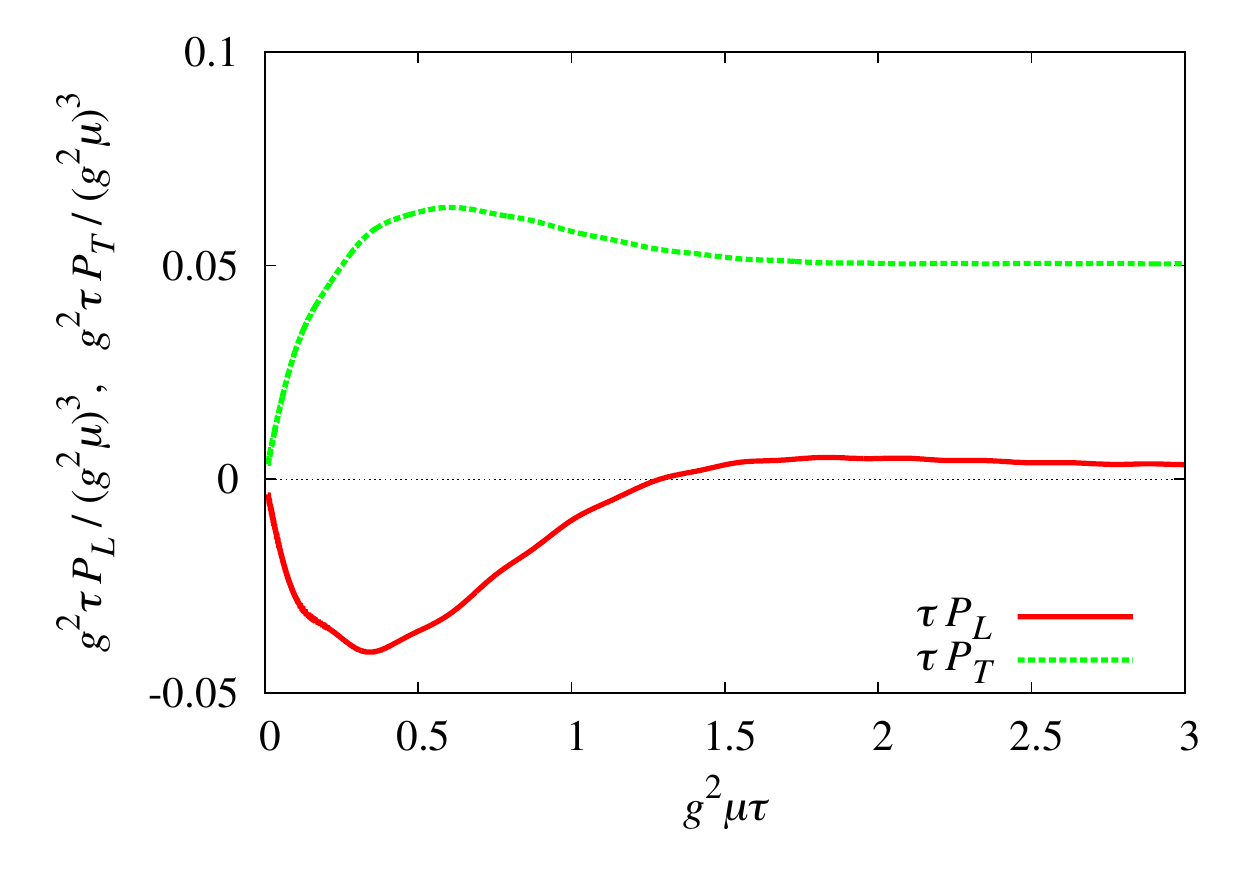}
 \end{center}
 \caption{Typical evolution of the color electric and magnetic fields
   in the MV model (left) and the transverse and the longitudinal
   pressure (right).  Figures are taken from
   \cite{Fukushima:2011nq}.}
 \label{fig:P}
\end{figure}
%---   figure   ---%

We can numerically solve the equations of motion in \eref{eq:eom} and
\eref{eq:eom2} on lattice discretized spacetime.  It is not mandatory
to use the link variables for classical theories, but the conventional
lattice formulation in terms of the link variables $U_\mu$ is
convenient to stabilize long time simulations.  It is then a bit
cumbersome to rewrite the initial condition \eref{eq:initial} in terms
of $U_\mu$, which was done in \cite{Krasnitz:1998ns}.

Ideally the results have no dependence on the choice of $\mux$ once all
variables are made dimensionless in the unit of $\mux$.  In the actual
calculation, however, this is not the case since the color average in
\eref{eq:average} is ultraviolet (UV) and infrared (IR) singular and
so the results depend on the lattice spacing $a$ and the system volume
$L^3$.  Nevertheless, such unphysical UV and IR sensitivity becomes
harmlessly mild once $\tau$ gets larger than $1/a$
\cite{Kovchegov:2005ss,Lappi:2006hq,Fukushima:2007ja} (and this is why
a naive expansion in terms of $\tau$ as attempted in
\cite{Fries:2006pv} completely fails due to singular $\tau/a$ terms;
see \cite{Fujii:2008km} for more details).  \Fref{fig:P} shows a
typical example of the temporal profile of the color electric and
magnetic fields (left figure) and also the transverse and the
longitudinal pressure (right figure) from the MV model simulation.
All physical quantities are made dimensionless in the unit of
$g^2\mux \sim \Qs$.  From these figures we see that the longitudinal
pressure $\PL$ goes negatively at first and approaches zero back for
$g^2\mux\tau\gtrsim 1$.  The longitudinal expansion
with $\PL=0$ means free-streaming, which will be closely discussed
later in \sref{sec:isotropization}.  To summarize the essential
features of the glasma initial condition, the color fields are boost
invariant ($\eta$ independent) and the longitudinal pressure is
negative.  We need to find some mechanism of violating the boost
invariance to decohere fields and to make $\PL$ turn back to positive.
For this purpose it is indispensable to consider quantum fluctuations
properly beyond the classical approximation.

%%%%%   CGC-based Scenarios for Thermalization  %%%%%
\subsection{CGC-based Scenarios for Thermalization}
\label{sec:bottomup}
The early-time dynamics in the heavy-ion collisions has a unique scale
$\Qs$, so that the proper unit to measure the time is
$\Qs^{-1}\sim 0.2\fm/c$ for RHIC and $\sim 0.1\fm/c$ for LHC.\ \ An
interesting and challenging question is whether we can somehow give an
analytical estimate for the thermalization time in terms of
$\alphas=g^2/(4\pi)$ and $\Qs$ for sufficiently small coupling
$\alphas\ll 1$.  This program was first addressed in
\cite{Baier:2000sb} (see also \cite{Kovchegov:2005ss} for a rather
negative conclusion) and the so-called ``bottom-up scenario'' has
grown popular.  Since this picture of the bottom-up thermalization
contains important view points for subsequent developments (as partly
seen in discussions in \sref{sec:thermal}), let us start this
subsection with a review of the bottom-up thermalization.

%%%   Bottom-up ...   %%%
\subsubsection{Bottom-up thermalization scenario}
The conclusion from the bottom-up thermalization scenario
\cite{Baier:2000sb} is that the parametrical expressions of the
thermalization time scale and the maximal temperature are,
respectively,
\begin{equation}
  \Qs\tau_{\rm th}\sim \alphas^{-13/5}\;,\qquad
  T/\Qs\sim \alphas^{2/5}\;.
\label{eq:thermtime_bottom}
\end{equation}
To understand these results, we should
first make it clear how to define thermalization.

In this scenario hard gluons with momenta $\sim \Qs$ are initially
produced and the thermalization time of soft gluons with momenta
$\sim T$ is defined when the energy of hard gluons is transferred to
soft gluons.  Let us first consider a branching process from a hard
gluon into gluons with a softer momentum $k_{\rm br}$.  If there are
$N_{\rm soft}$ soft gluons and their population is a thermalized one by
$N_{\rm soft}\sim T^3$, as we explain soon later, the following
relations can be shown:
\begin{equation}
 k_{\rm br} \sim \alpha_s^4 T^3 \tau_{\rm th}^2\;,\qquad
 T \sim \alphas^3 \Qs^2 \tau\;.
\label{eq:branch}
\end{equation}
Then, once these are accepted, the energy flow from hard to soft
gluons should be terminated when $k_{\rm br}\sim \Qs$, which, together
with $T\sim \alphas^3\Qs^2\tau_{\rm th}$, means that
$k_{\rm br}\sim \alphas^{13}\Qs^6 \tau_{\rm th}^5 \sim \Qs$ leading
immediately to the thermalization time scale and the initial
temperature at $\tau=\tau_{\rm th}$ as given in
\eref{eq:thermtime_bottom}.

%---   figure   ---%
\begin{figure}
 \begin{center}
 \includegraphics[width=0.4\textwidth]{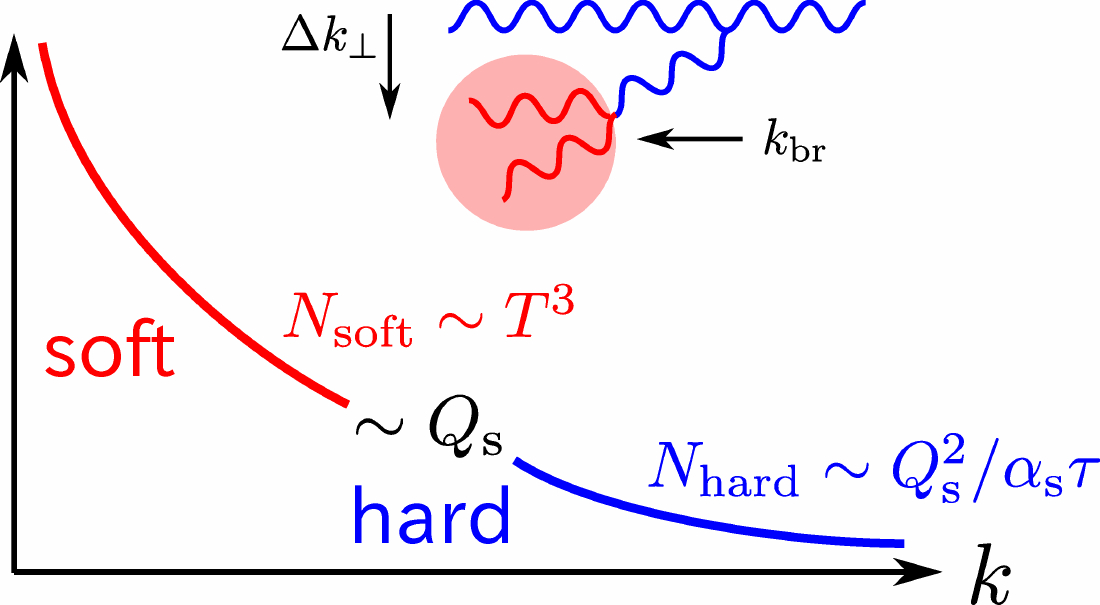}
 \end{center}
 \caption{Schematic illustration of the bottom-up thermalization
   scenario.}
 \label{fig:bottomup}
\end{figure}
%---   figure   ---%

To understand the first relation in \eref{eq:branch} let us consider a
formation time $\tau_{\rm f}$ needed for one emission process, which
is estimated by the uncertainty principle as
\begin{equation}
 \tau_{\rm f} \sim \frac{1}{k^+}
             \sim \frac{k_{\rm br}}{\Delta k_\perp^2}\;.
\end{equation}
Then, the energy is deposited to the thermal medium by further hard
splitting processes and the time taken by these processes defines the
thermalization time $\tau_{\rm th}$.  Parametrically,
$\tau_{\rm f}=\alphas \tau_{\rm th}\ll \tau_{\rm th}$.  Now, we need to
know what $\Delta k_\perp$ is, which reflects the thermal properties
in the soft sector.  Using a diffusion constant
$\hat{q}\equiv \rmd\langle k_\perp^2\rangle/\rmd t$, it is obvious
that $\Delta k_\perp^2 \sim \hat{q}\cdot \tau_{\rm f}$, and $\hat{q}$
can be parametrized as $\hat{q}\sim m_{\rm D}^2/\lambda$ with the
Debye mass $m_{\rm D}$ and the mean-free path $\lambda$.  In a thermal
medium $m_{\rm D}^2\sim \alphas T^2$ and $\lambda\sim 1/(\alphas T)$,
which eventually yields $\hat{q}\sim \alphas^2 T^3$.  Therefore, we
can have a relation:
\begin{equation}
 \Delta k_\perp^2 \sim \frac{k_{\rm br}}{\tau_{\rm f}}
  \sim \alphas^2 T^3 \tau_{\rm f} \;.
\end{equation}
The first expression in \eref{eq:branch} is a result from plugging
$\tau_{\rm f}=\alphas \tau_{\rm th}$ into the above.

The second in \eref{eq:branch} originates from the energy balance.  In
terms of the rate of the gluon production, $\rmd N/\rmd\tau$, the
energy flow per unit time should be $k_{\rm br}\cdot \rmd N/\rmd\tau$
that is equated to an increase in thermal energy by
$\rmd(T^4)/\rmd\tau$.  Because softer gluons are emitted from a hard
gluon whose density is $\sim \Qs^3/(\alphas \Qs\tau)$ (where the gluon
distribution function in the saturation regime is $\sim 1/\alphas$ and
$\Qs\tau$ in the denominator represents the longitudinal expansion
effect), the rate should be characterized as
$\rmd N/\rmd\tau\sim Q_s^2/(\alpha_s \tau^2)$.  Therefore,
$\rmd(T^4)/\rmd\tau\sim T^3\rmd T/\rmd\tau\sim
(\alpha_s^4 T^3\tau^2)\cdot Q_s^2/(\alpha_s \tau^2)$, which concludes
that the temperature grows up linearly as expressed in the second
relation in \eref{eq:branch}.  As discussed in the original work
\cite{Baier:2000sb} the above-mentioned qualitative derivations can be
more quantified by means of the Boltzmann equation.  The Boltzmann
equation is actually a very useful tool and is widely used for other
scenarios like a CGC-driven BEC, as introduced in details in
\sref{sec:thermal}.

%%%   Plasma and glasma instabilities   %%%
\subsubsection{Plasma and glasma instabilities}
The prefactor of $\tau_{\rm th}$ from the bottom-up thermalization
scenario is expected to be not much different from the unity.  If we
take the parametric estimate literally, $\alphas^{-13/5}\sim 23$ for
$\alphas\sim 0.3$ and it is difficult to account for thermalization
within a reasonable time scale, namely, a few times $\Qs^{-1}$ or even
earlier.  There must be some missing mechanism that should accelerate
thermalization.

\begin{figure}
 \begin{center}
 \includegraphics[width=0.5\textwidth]{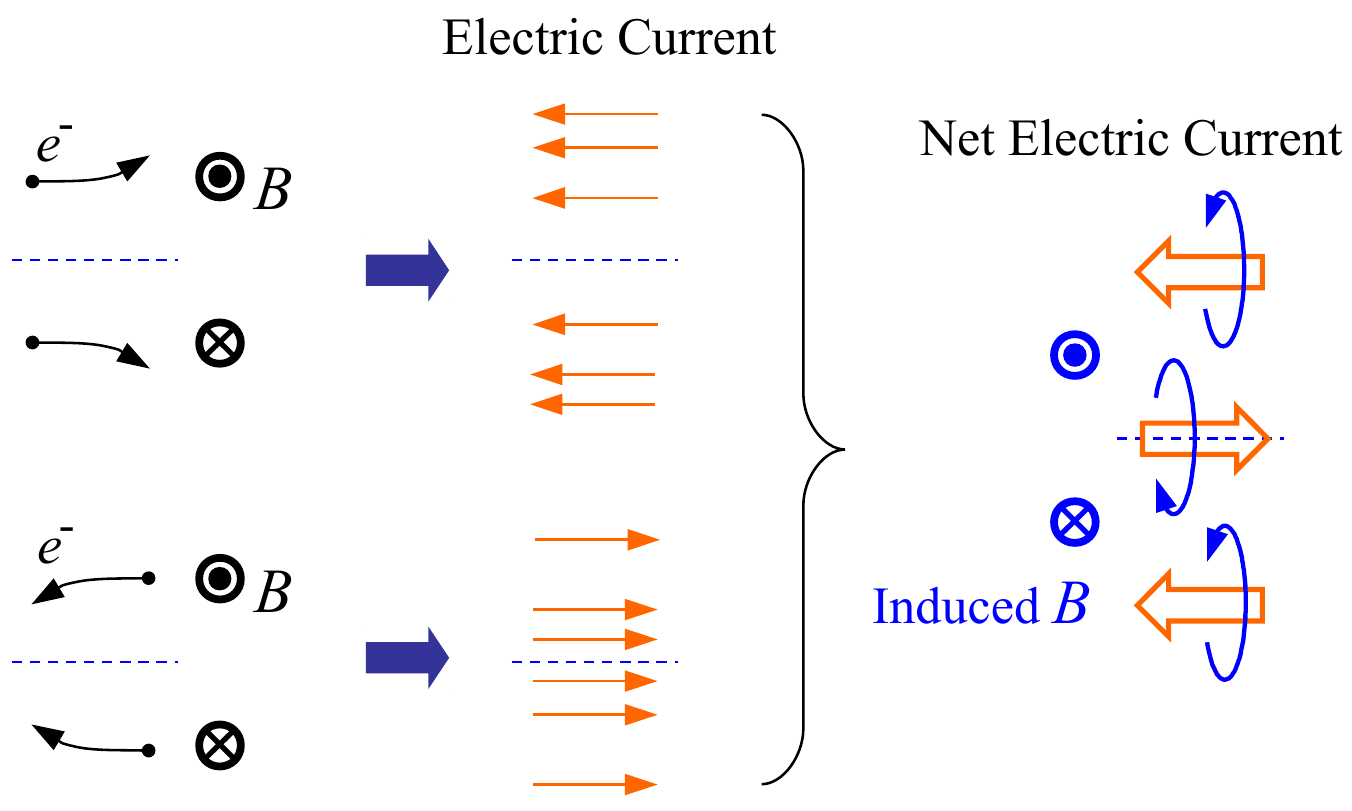}
 \end{center}
 \caption{Schematic illustration of the Weibel instability.  Spatial
 inhomogeneity of the magnetic field is strengthened by the
 backreaction of electron motions under the magnetic field.}
 \label{fig:weibel}
\end{figure}

It has been pointed out that a plasma in general has various
instabilities and so the isotropization can be quickly driven by QCD
counterparts of them, namely, \textit{QCD plasma instabilities}
\cite{Mrowczynski:1988dz};
especially, an instability induced by strong anisotropy in the
momentum distribution is important \cite{Mrowczynski:1993qm} (see
\cite{Arnold:2003rq} for comprehensive and analytical arguments on QCD
plasma instabilities).  Among several instabilities, it is believed
that the Weibel instability is the most relevant for the heavy-ion
collisions that spontaneously forms a filamentation pattern.  It would
be instructive to take a look at the Weibel instability in a QED
plasma with the electric current and the magnetic field $B$.
\Fref{fig:weibel} captures the
essential idea of the Weibel instability.  Suppose that there is
spatial inhomogeneity in $B$, electron motions are affected by the
magnetic field.  The upper situation in the figure shows the electron
motion in one direction, and the lower in the opposite direction.  In
both cases the same pattern of the electric current appears as
depicted in the right of the figure.  Induced magnetic fields are
sourced by these electric currents and new $B$ turns out to strengthen
the initial spatial inhomogeneity in $B$.  Since the initial
disturbance is amplified each time the backreaction from electron
motions is taken into account, the filamentation pattern grows up
exponentially fast that signals for an instability.

%---   figure   ---%
%---   figure   ---%

In the pure Yang-Mills theory the system has no direct counterpart of
electrons, i.e., (approximately) no quarks in the initial dynamics,
and yet, gluons are color charged particles.  Therefore, color
magnetic fields at soft scale and color charged gluons at hard scale
are sufficient ingredient for the realization of the non-Abelian
Weibel instability.  Let us recall that the CGC initial condition is
boost invariant, and $\PL$ is negative as long as boost-invariant
color flux tubes are extending between two nuclei.  Thus, it is
indispensable to violate boost invariance by breaching color flux
tubes with quantum fluctuations.  Actually, this should be physically
interpreted as particle production due to string breaking processes.
Because we are interested in the fate of boost-invariant background
fields $\calA^\mu$, it would be convenient to introduce a Fourier
transform as
\begin{equation}
 \dcalA^\mu(\tau,\nu,\bkt) \equiv \int \rmd\eta\, \rme^{-\rmi \nu\eta}
  \calA^\mu(\tau,\eta,\bkt)\;.
\label{eq:nu}
\end{equation}
The physical meaning of $\nu$ is a dimensionless wave-number to
quantify inhomogeneity along the longitudinal beam axis.  Fields at
$\nu=0$ represent boost-invariant backgrounds, and the definition
gives a relation; $\nu=t k_z + z k_0$ (in this review, we do not
distinguish covariant and contravariant vectors; $k_z=k^z$ simply,
except for the notations for the light-cone and the Bjorken
coordinates).  Then, using this Fourier transformed variables, we can
write the classical gluon fields as
$2\pi\delta(\nu)\calA^\mu(\tau,\bkt)+\dcalA^\mu(\tau,\nu,\bkt)$ with
boost-invariant CGC fields and instability-driven fluctuations.  As
long as the latter is smaller than the former, we can investigate the
instability using linearized equations of motion; i.e., in the Bjorken
coordinates, the transverse fields should satisfy:
\begin{equation}
 \partial_\tau \tau \partial_\tau \dcalA_i
  = -\frac{\nu^2}{\tau}\dcalA_i
    +\tau \mathcal{G}^{-1}_{ij}[\calA] \dcalA_j\;.
\label{eq:eom_inst}
\end{equation}
with the full gluon propagator $\mathcal{G}^{\mu\nu}[\calA]$ in the
presence of background $\calA^\mu$.  If $\mathcal{G}^{-1}_{ij}$ has a
positive eigenvalue $\lambda$, then the solution of the above equation
should generally take the following form \cite{Fukushima:2007ja}:
\begin{equation}
 \dcalA_i \sim c_1 \Re I_{\rmi\nu}(\sqrt{\lambda}\tau)
              +c_2 \Im I_{\rmi\nu}(\sqrt{\lambda}\tau)\;,
\end{equation}
where $I_{\rmi\nu}(\sqrt{\lambda}\tau)$ is the modified Bessel
function.  Given some initial condition, the evolution of
$\dcalA^\mu(\tau)$ is deterministic, and its time dependence should be
exponential if $c_1\neq 0$ in the above.

%---   figure   ---%
\begin{figure}
 \begin{center}
 \includegraphics[width=0.45\textwidth]{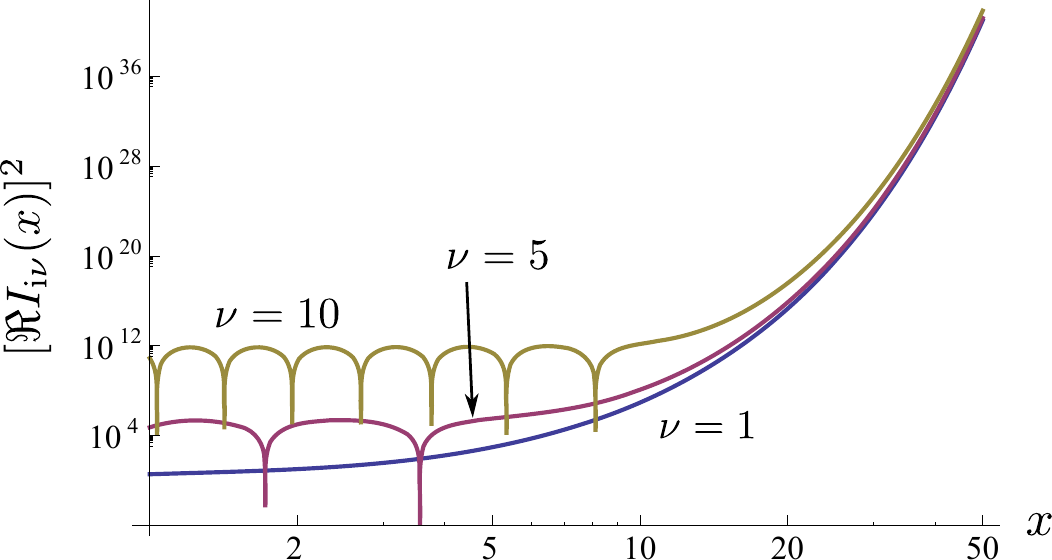} \hspace{1em}
 \includegraphics[width=0.45\textwidth]{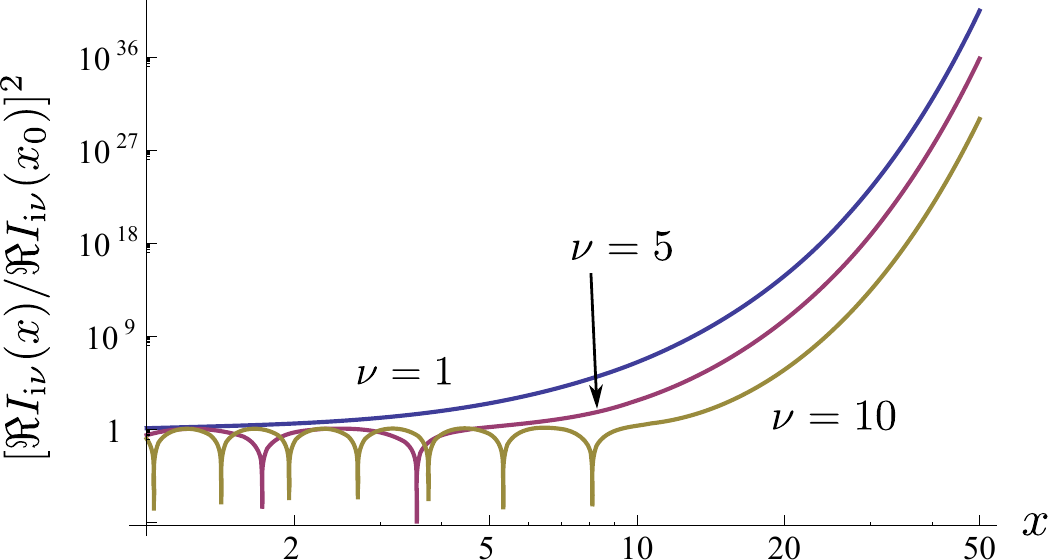}
 \end{center}
 \caption{Typical behavior of unstable modes in one-dimensionally
   expanding systems.  (Left) Squared quantities of the modified
   Bessel functions for $\nu=1$, $5$, and $10$.  (Right) Squared
   quantities of the modified Bessel functions normalized at
   $x_0=0.1$.}
 \label{fig:bessel}
\end{figure}
%---   figure   ---%

Let us take an even closer look at $\Re I_{\rmi\nu}(x)$ to have an
intuitive feeling about instabilities in one-dimensional expanding
systems.  \Fref{fig:bessel} plots $[\Re I_{\rmi\nu}(x)]^2$ in two
different ways.  The left figure shows the values of the functions as
they are for $\nu=1$, $5$, and $10$.  We see that the oscillatory
behavior lasts longer for larger $\nu$, which can be easily explained
from \eref{eq:eom}.  As long as the first term enhanced by
$\nu^2/\tau$ overwhelms the right-hand side, no unstable behavior
appears.  This observation fits in with our intuition: the expansion
tends to inhibit instability.  The oscillatory region itself, however,
does not delay the onset of instability as is the case in the left of
\fref{fig:bessel}, showing asymptotic convergence to the same curve.
In most physical cases the weight for larger-$\nu$ modes should be
more suppressed (otherwise, $\Re I_{\rmi \nu}(x)$ becomes unphysically
large as $\nu$ increases), and to see this effect, the right of
\fref{fig:bessel} shows
$[\Re I_{\rmi\nu}(x)/\Re I_{\rmi\nu}(x_0=0.1)]^2$, so that different
$\nu$ modes are all normalized at $x=x_0$.  Then, for larger $\nu$,
the weight is smaller and the waiting time for instability becomes
larger \cite{Fukushima:2007ja,Fujii:2008dd}.  Because such delaying
effects are so sensitive to $\nu$-dependent weight, we must know
the spectrum of initial fluctuations very precisely to locate the
onset of instabilities and to account for early isotropization
quantitatively.

Now let us return to discussions on the QCD Weibel instability.  There
are many analytical and numerical studies on the Weibel instability
and it is not realistic to try to cover all of them here.  As a
typical and comprehensible example, let us pick up one fairly
analytical formulation in \cite{Rebhan:2004ur} (see also related
numerical works \cite{Dumitru:2005gp,Dumitru:2006pz}).  The basic
setup is a combination of the Yang-Mills equation and the Vlasov
equation (or Wong's equation in \cite{Dumitru:2005gp}).  The color
fields represent the soft components of gluons as is naturally
implemented in the CGC theory.  The hard components are split into the
color-neutral part and the colored part.  The neutral part is assumed
to have anisotropic distribution,
\begin{equation}
  f_0(\bk) = f_{\rm iso}(\sqrt{\bkt^2+k_\eta^2/\tau_{\rm iso}^2})\;,
\label{eq:anisotropic}
\end{equation}
where $f_{\rm iso}(\bk)$ is an isotropic distribution function.
The colored part, $\delta f^a(\bk,x)$, that represents a counterpart
of electrons in \fref{fig:weibel}, should be then determined by the
Vlasov equation that reads (with the collision term neglected):
\begin{equation}
 k_\alpha \calD^{\alpha ab} \delta f^b(\bkt,k_\eta)
  = g k_\alpha \calF^{\alpha\beta a} \frac{\partial}{\partial p^\beta}
  f_0(\bkt,k_\eta) \;,
\end{equation}
in the Bjorken coordinates.  Once $\delta f^a(\bkt,k_\eta)$ is solved,
the color fields should satisfy the Yang-Mills equations with a color
source provided by $\delta f^a(\bkt,k_\eta)$, that is,
\begin{equation}
 \frac{1}{\tau} \calD_\alpha^{ab} (\tau\calF^{\alpha\beta b})
  = j^{\beta a} = \frac{g}{2}\int\frac{\rmd^2\bkt\,\rmd y}{(2\pi)^3}
  \,k^\beta\,\delta f^a(\bkt,k_\eta)\;,
\end{equation}
where $y\equiv \mathrm{atanh}(k_0/k_z)$.  The information on the
isotropic distribution is totally encompassed in the Debye mass, which
is defined in terms of the distribution function as
\begin{equation}
  \mD^2 \equiv g^2\int_0^\infty \frac{\rmd^3\bk}{(2\pi)^3 2\omega(\bk)}\,
  f_{\rm iso}(\bk)\;.
\label{eq:Debye}
\end{equation}
It must be noted that $f_{\rm iso}(\bk)$ in the above is spin-summed
and color-averaged one;  if the distribution function is the one per
spin and color, the definition of $\mD^2$ should be multiplied by
$2C_A=2\Nc$, which more often appears in the literature.  The
linearized Yang-Mills equations after eliminating
$\delta f^a(\bkt,k_\eta)$ should dictate the temporal evolution of
fluctuation modes and, for late time $\tau\gg \tau_0\gg \tau_{\rm iso}$
where $\tau_0$ represents the initial time, the Yang-Mills equations
are reduced to
\begin{eqnarray}
 && \bigl[ \partial_\tau^2 \tau \partial_\tau \tau \partial_\tau
  +\nu^2\partial_\tau^2 + \mu\partial_\tau^2 \tau - \mu\nu^2 \tau^{-1}
  \bigr]\dcalA^i(\tau,\nu) = 0\;,
\label{eq:tAi} \\
 && \bigl[ \partial_\tau \tau^{-1}\partial_\tau + 2\mu \tau^{-2}\bigr]
  \dcalA_\eta(\tau,\nu) = 0\;,
\end{eqnarray}
where $\mu\equiv \frac{\pi}{8}m_{\rm D}^2 \tau_{\rm iso}$, which
actually represent the concrete contents of \eref{eq:eom}.  These are
easily solvable using the (modified) Bessel functions.  In fact,
$\dcalA_\eta(\tau,\nu)$ can be given by a linear superposition of
oscillatory Bessel functions and thus it is concluded that
$\dcalA_\eta(\tau,\nu)$ has no instability in the linear regime.  It
is found that, even for $\nu\gg 1$ (which is less unstable according
to the previous discussions) $\dcalA^i(\tau,\nu)$ is a linear
superposition of modified Bessel functions as
\begin{equation}
 \dcalA^i(\tau,\nu\gg 1) \simeq c_1\, \sqrt{\tau} I_1(2\sqrt{\mu\tau})
  + c_2\, \sqrt{\tau} K_1(2\sqrt{\mu\tau})\;,
\end{equation}
which is an exponentially growing function and this diverging behavior
manifests the non-Abelian Weibel instability.  The appearance of
$\sim I_n(c\sqrt{\mu\tau}) \sim (\mu\tau)^{-1/4}\rme^{c\sqrt{\mu\tau}}$ 
is typical in one-dimensional expanding systems;  the exponential
growth is not like $\sim \rme^{c\tau}$ but $\sim \rme^{c\sqrt{\mu\tau}}$
due to expansion.

In the approach with the Vlasov and the Yang-Mills equations the
separation between the field of soft gluons and the particle of hard
gluons seems to be an artificial choice.  In principle, the physical
results should not depend on where the separation scale is, but it is
quite non-trivial to verify this (see \cite{Dumitru:2006pz} for an
example of explicit check).  It would be desirable to build a simple
and unifying description to deal with both soft and hard components
within a common framework.  Promising results actually came out from
the glasma simulation with initial fluctuations incorporated.

\begin{figure}
 \begin{center}
 \includegraphics[width=0.6\textwidth]{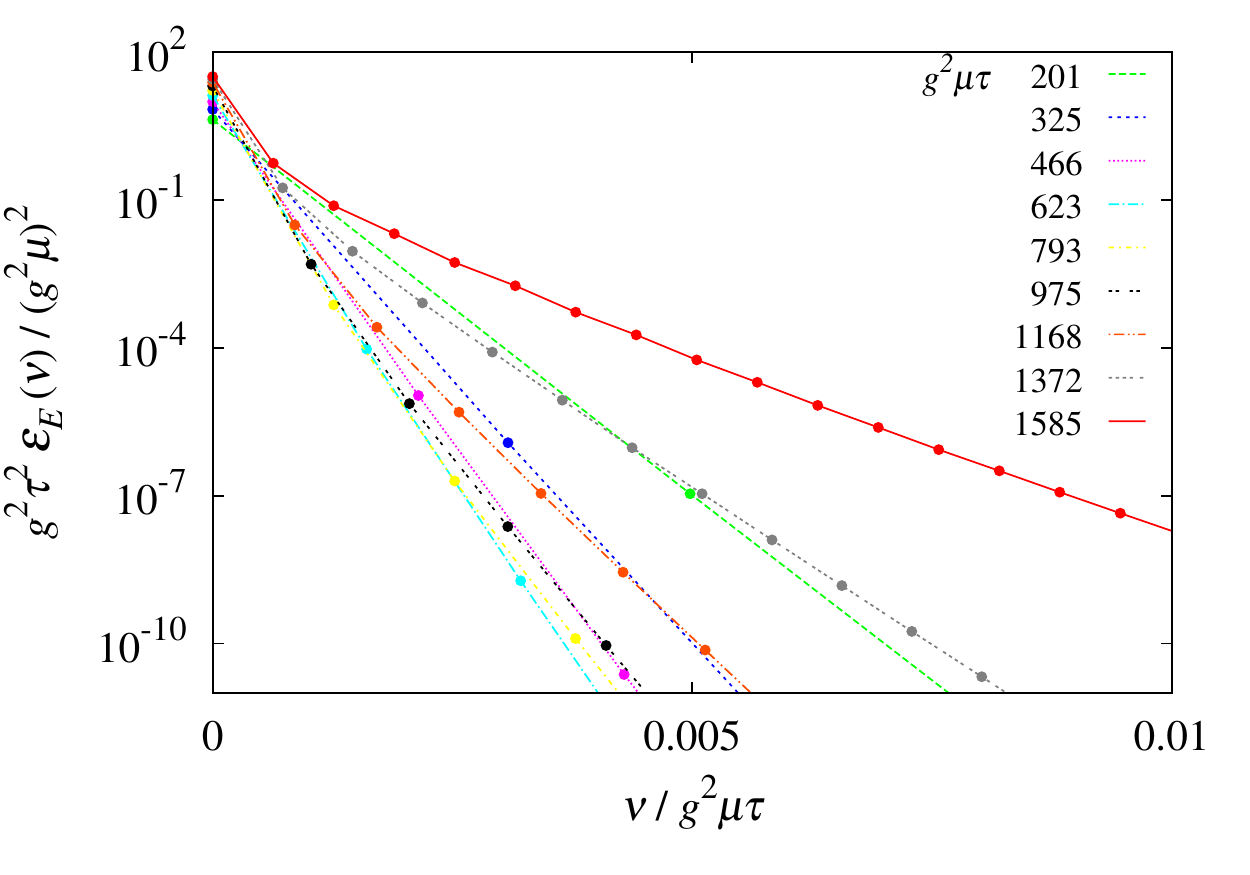}
 \end{center}
 \caption{Spectral decomposition of the CGC electric energy density as
   a function of wave-number $\nu$ for various time $\tau$.  Initially
   all the energy is stored only at $\nu=0$, which diffuses to larger
   $\nu$ as the time goes.  Figure is taken from
   \cite{Fukushima:2011nq}.}
 \label{fig:glasma_inst}
\end{figure}

Instead of solving coupled equations for $\calA_\mu$ and
$\delta f^a(\bk)$ and integrating $\delta f^a(\bk)$ out, we can
directly write down a counterpart of \eref{eq:eom} by linearizing the
classical Yang-Mills equations with
$2\pi\delta(\nu)\calA^\mu(\tau,\bk_\perp)+\dcalA^\mu(\tau,\nu,\bk_\perp)$.
In fact, we do not have to perform the linearization, but we can just
solve the full classical Yang-Mills equations with initial fluctuation
seeds $\dcalA^\mu(\tau,\nu,\bk_\perp)$ to break boost invariance.  In
this way, unstable behavior has been discovered, which is referred to
as \textit{glasma instability} \cite{Romatschke:2005pm}.  Physically
speaking, by construction, the origin of the glasma instability should
be the same as that of the plasma (Weibel) instability as emphasized
in \cite{Romatschke:2006nk}, but there is no clear correspondence
between the glasma and the plasma instabilities on the algebraic
level.  In a sense, as numerically observed in \fref{fig:glasma_inst}
for example, the glasma instability could be understood as a diffusion
process in $\nu$ space from the CGC field at $\nu=0$ to higher $\nu$
modes, which was investigated by mode-by-mode analysis in
\cite{Fukushima:2011nq}, and a sort of avalanche behavior was verified
in \cite{Fukushima:2013dma} (see also \cite{Dumitru:2006pz} in which
the UV avalanche was first pointed out).  This is an example of the
self-similarity and the spectral cascade phenomenon that will be more
discussed in \sref{sec:isotropization} and \sref{sec:thermal}.  We
also note that the classical Yang-Mills equations are highly
non-linear, and so there may be instabilities associated with chaotic
behavior of solutions \cite{Biro:1993qc}.  In fact, in some numerical
simulations \cite{Heinz:1996wx,Kunihiro:2010tg}, the Lyapunov
exponents have been extracted, which is useful for the computation of
the Kolmogorov-Sinai entropy \cite{Iida:2013qwa}.

%---   figure   ---%
%---   figure   ---%

%%%%%   Real-time Formulations   %%%%%
\subsection{Real-time Formulations}
\label{sec:formulations}
We have already previewed some results from the kinetic equation and
the classical field.  The former is effective for a regime where the
gluon distribution is dilute.  Once the gluon amplitude reaches the
saturation regime, the expansion of the collision term with respect to
the gluon distribution does not work, and the classical approximation
makes better sense.  The semi-classical method has been highly
sophisticated into a form of the classical statistical approximation
nowadays, which, however, may suffer the UV singularities.  Finally,
we will quickly look over some other methods.

%%%   Dilute regime --- kinetic equation   %%%
\subsubsection{Dilute regime --- kinetic equation}
\label{sec:dilute}
Let us begin with a classical example of simple scalar field theory.
In the dilute regime at weak coupling, the Boltzmann equation should
be an appropriate description of the real-time dynamics.  For the
distribution function $f(\bk,\bx,t)$ the scalar Boltzmann equation
reads:
\begin{equation}
 \frac{\rmd f}{\rmd t} =
 \biggl( \frac{\partial}{\partial t}
         + \dot{\bx}\frac{\partial}{\partial\bx}
         + \dot{\bk}\frac{\partial}{\partial\bk} \biggr)
  f(\bk,\bx,t) = \biggl(\frac{\partial f}{\partial t}
  \biggr)_{\rm coll}\;,
\label{eq:Boltzmann}
\end{equation}
where the last term represents the collision term, which can be
diagrammatically calculated at weak coupling.  The simplest example is
an elastic $2\leftrightarrow 2$ process, for which the collision term
should take a conventional expression,
\begin{eqnarray}
\fl
\biggl(\frac{\partial f(\bk_1)}{\partial t}
\biggr)_{2\leftrightarrow 2} &=& \frac{1}{4\nu\, \omega(\bk_1)}
  \int_{\bk_2,\bk_3,\bk_4}
  (2\pi)^4 \delta^{(4)}(k_1+k_2-k_3-k_4)\,|\calM_{2\leftrightarrow 2}(\bk)|^2
  \nonumber\\
\fl
&& \quad \times \Bigl\{f_3 f_4 (1+f_1)(1+f_2)
  - f_1 f_2 (1+f_3)(1+f_4) \Bigr\}\;.
\label{eq:twotwo}
\end{eqnarray}
Here we introduced a compact notation;
$\int_{\bk}\equiv \int\,\rmd^3\bk/[2\omega(\bk)(2\pi)^3]$ with
$\omega(\bk)=|\bk|$ for massless bosons, $f_i \equiv f(\bk_i)$, and
$\nu$ represents the degeneracy factor associated with internal
quantum number (such as spin degeneracy).  When the system gets
equilibrated, $\rmd f/\rmd t=0$ and so the detailed balance is
realized, from which the Bose-Einstein distribution function is
derived as follows.  Let us require that
$f_3f_4(1+f_1)(1+f_2)-f_1f_2(1+f_3)(1+f_4)=0$ for
arbitrary $\bk_1$, $\bk_2$, $\bk_3$, and $\bk_4$, which is a
sufficient (but not necessary) condition to let the collision term
vanish.  Then, by taking logarithms, we can show,
\begin{equation}
   \ln\biggl(\frac{1+f_1}{f_1}\biggr)
 + \ln\biggl(\frac{1+f_2}{f_2}\biggr)
 = \ln\biggl(\frac{1+f_3}{f_3}\biggr)
 + \ln\biggl(\frac{1+f_4}{f_4}\biggr) \;.
\end{equation}
This means that $\ln[(1+f)/f]$ should be a conserved quantity, and so
should be expressed as a linear combination of basic conserved
quantities; $1$, $\bk$, and $k^0=\omega(\bk)$ as
\begin{equation}
 \ln\biggl[\frac{1+f(\bk)}{f(\bk)}\biggr] = \frac{\mu}{T}
  - \frac{k^\mu u_\mu}{T}\quad\Rightarrow
 \quad f(\bk)=\frac{1}{\rme^{(k^\mu u_\mu - \mu)/T}-1} \;.
\end{equation}
Here, $u_\mu$ represents the fluid four-velocity.

The thermalization problem of the QGP or the isotropization in modern
language was first investigated in \cite{Baym:1984np} using the
Boltzmann equation with the relaxation time approximation (RTA).
Because the calculations are quite instructive to explain the basic
features of the QGP, below, we shall reiterate the main steps of
calculations in \cite{Baym:1984np}.  In the heavy-ion collision the
system cannot be homogeneous in the Cartesian coordinates because the
longitudinal velocity is $u=z/t$ and the Lorentz time dilatation
becomes greater for larger $z$.  So, we should keep
$\partial/\partial z$ in the left-hand side of \eref{eq:Boltzmann},
while $\dot{\bk}$ drops off without external force.  Then, once boost
invariance is imposed, $z$ dependence is uniquely fixed through
$k_z(z)=\gamma(k_z - k_0 u)$ where $k_z$ and $k_0$ are the
longitudinal momentum and the energy at $z=0$.  From this, it is easy
to see that $\partial/\partial z=-(k_0/t)\partial/\partial k_z$ if
they act on a function of $k_z(z)$.  Therefore, in this case of
boost-invariant expansion, the Boltzmann equation takes a form of
\begin{equation}
 \biggl( \frac{\partial}{\partial t} - \frac{k_z}{t}
  \frac{\partial}{\partial k_z} \biggr) f(\bkt,k_z,t)
  = \biggl(\frac{\partial f}{\partial t}\biggr)_{\rm coll}\;,
\label{eq:Boltzmann_Bj}
\end{equation}
apart from transverse dynamics that we neglect.  Below we drop $z$
from $f(\bkt,k_z,z,t)$ focusing on the mid rapidity region only.  In
the absence of collision, we have the free-streaming solution; i.e.,
$f(\bkt,k_z,t)=f(\bkt,k_z t/t_0)$ solves \eref{eq:Boltzmann_Bj}.  In
the free-streaming case, the local energy density becomes,
\begin{equation}
\fl\qquad\qquad
 \varepsilon(t) = \int\frac{\rmd^3\bk}{(2\pi)^3} |\bk|\,
  f(\bk,t)
 = \frac{t_0}{t}\int\frac{\rmd^2\bkt \rmd\xi}{(2\pi)^3}
  \sqrt{\bkt^2 + \xi^2}\, f_0(\bkt,\xi)\; \propto\; \frac{1}{t}\;,
\label{eq:free}
\end{equation}
which is a natural consequence from one-dimensional expansion.  In the
RTA in which the analytical calculation is feasible, the collision
term is assume to be as simple as
\begin{equation}
 \biggl(\frac{\partial f}{\partial t}\biggr)_{\rm coll}
  = -\frac{1}{\tau_{\rm rel}} \bigl[ f(\bk_\perp,k_z,t)
    - f_{\rm eq}(\bk,t) \bigr]\;,
\label{eq:RTA}
\end{equation}
where $\tau_{\rm rel}$ represents the relaxation time and
$f_{\rm eq}(\bk,t)$ is the Bose-Einstein distribution function at the
temperature $T(t)$.  Generally speaking, $\tau_{\rm rel}$ is a
function of time and momenta, but if we adopt a constant
$\tau_{\rm rel}$, we can analytically solve the Boltzmann equation
under the initial condition of $f(\bkt,k_z,t=t_0)=f_0(\bkt,k_z)$ as
\begin{equation}
\fl
 f(\bkt,k_z,t) = \rme^{(t_0-t)/\tau_{\rm rel}} f_0(\bkt,k_z t/t_0)
 + \int_{t_0}^t \frac{\rmd t'}{\tau_{\rm rel}}\,
 \rme^{(t'-t)/\tau_{\rm rel}}\,
  f_{\rm eq}(\sqrt{\bkt^2+(k_z t/t')^2},t') \;.
\end{equation}
From this form of the solution, an integral equation for
$\varepsilon(t)$ can be derived \cite{Baym:1984np}, which can be
solved for $t\gg \tau_{\rm rel}$ (with the energy conservation and an
assumption that the initial distribution is peaked at $p^z=0$) leading
finally to
\begin{equation}
 \frac{\varepsilon(t)}{\varepsilon(t_0)} \simeq 1.22\;
  \frac{t_0 \tau_{\rm rel}^{1/3}}{t^{4/3}} \;\propto\; \frac{1}{t^{4/3}}\;.
\label{eq:Bjorken_scaling}
\end{equation}
This $t$ dependence makes a sharp contrast to the free-streaming one
in \eref{eq:free} and should be interpreted as the complete
isotropization.  Actually, the conservation equation in the expanding
system reads:
\begin{equation}
 \frac{\rmd \varepsilon}{\rmd \tau}
 + \frac{\varepsilon + \PL}{\tau} = 0\;,
\label{eq:e_PL}
\end{equation}
where $\PL$ is the longitudinal pressure as defined in \eref{eq:PL}.
If $\PL=0$ in the free-streaming case, $\varepsilon \sim 1/\tau$ as we
have already seen in \eref{eq:free}.  (It should be noted that we took
$z=0$ to simplify discussions, so that $\tau$ is just $t$ then.)  Once
$\PL=\PT$ is realized and the conformality is approximately realized
as $\varepsilon-2\PT-\PL\approx 0$, then $\PL\approx \varepsilon/3$
and we see that $\varepsilon\sim 1/\tau^{4/3}$ is concluded from
\eref{eq:e_PL}.  In summary, if the interaction is turned off, the
free-streaming solution leads to $\varepsilon \propto 1/\tau$, and if
the interaction is strong enough to achieve the complete
isotropization, the hydrodynamic scaling (in a sense of old-fashioned
characterization) follows as $\varepsilon \propto 1/\tau^{4/3}$.  In
reality it is quite unlikely that the system can be fully isotropized
in the heavy-ion collisions and the exponent should be something
between $-1$ and $-4/3$.  For the reliable determination of the
exponent, the RTA is a too crude approximation, and the most serious
obstacle is that the RTA would violate conservation laws, which may be
cured in the Lorentz model, but it should be of course much better if
the QCD interaction is systematically considered.

%---   figure   ---%
\begin{figure}
 \begin{center}
 \includegraphics[width=0.5\textwidth]{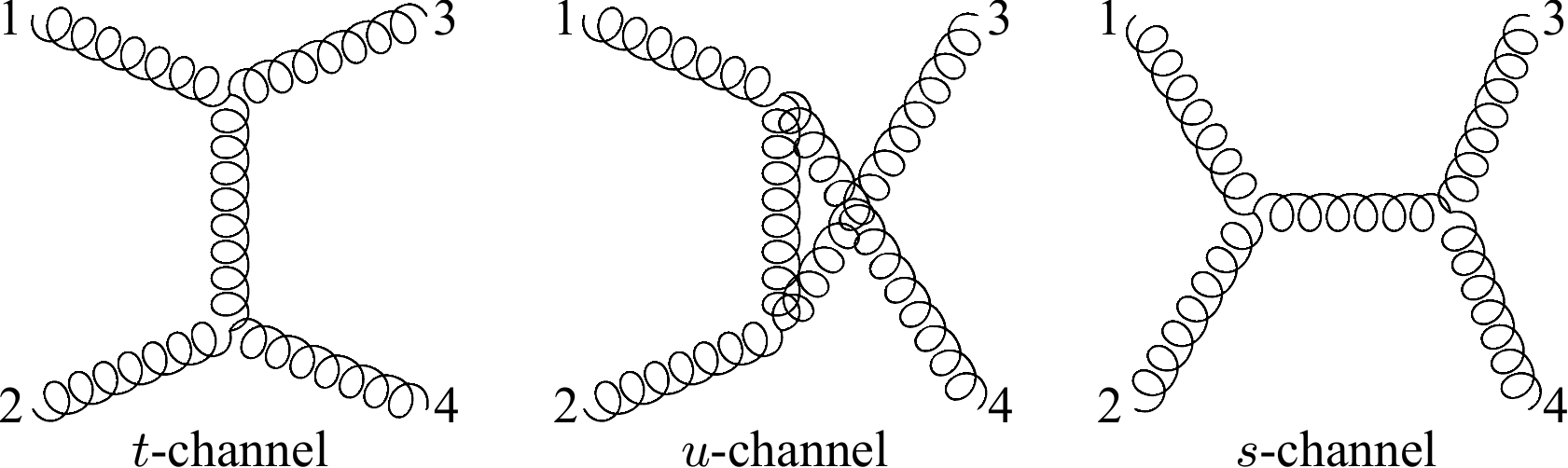} \vspace{1em}\\
 \includegraphics[width=0.7\textwidth]{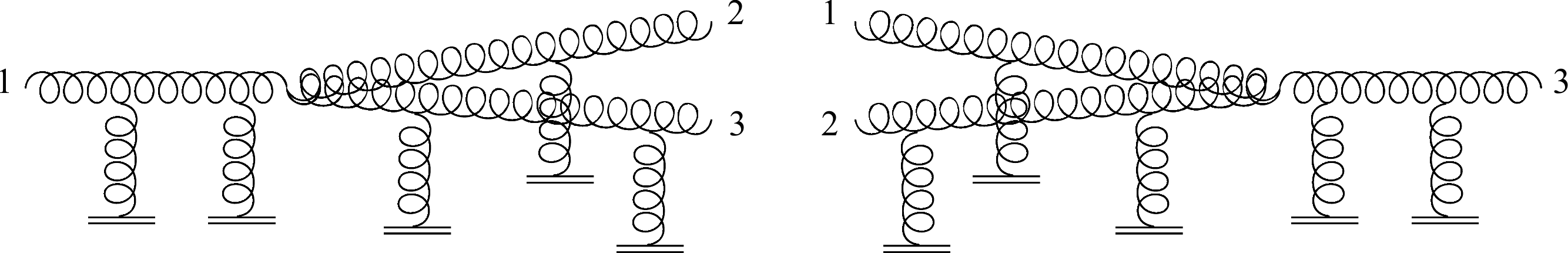}
 \end{center}
 \caption{The collisions terms of $2\leftrightarrow 2$ scattering
   (upper) and $1\leftrightarrow 2$ scattering (lower).  The reversed
   processes to increase $1$ are omitted.}
  \label{fig:feyn}
\end{figure}
%---   figure   ---%

For this purpose the effective kinetic theory (EKT) of QCD
\cite{Arnold:2002zm} has been developed, with which the shear
viscosity calculation is performed and also the jet energy loss is
evaluated.  The EKT consists of the Boltzmann equation with two
scattering terms;  an elastic $2\leftrightarrow 2$ scattering and an
inelastic $1\leftrightarrow 2$ scattering.  The former is easy to find
from the usual Feynman rule.  Because the triple-gluon vertex has one
derivative, for example, the $t$-channel scattering in the upper left
in \fref{fig:feyn} leads to
$|(k_1+k_3)\cdot(k_2+k_4)/(k_3-k_1)^2|=(s-u)^2/t^2=1-4us/t^2$ using
the Mandelstam variables, $s=(k_1+k_2)^2$, $t=(k_1-k_3)^2$, and
$u=(k_1-k_4)^2$ with four-vector notation.  Summing the $u$-channel
and $s$-channel contributions up together with the quartic-gluon
vertex term, the $2\leftrightarrow 2$ matrix element eventually
amounts to
\begin{equation}
  |\calM_{2\leftrightarrow 2}|^2 = 16g^4 d_A C_A^2 \biggl( 3 - \frac{us}{t^2}
  - \frac{st}{u^2} - \frac{tu}{s^2} \biggr)\;,
\label{eq:Mtwotwo}
\end{equation}
where $d_A=\Nc^2-1$ and $C_A=\Nc$.

In contrast to this, the $1\leftrightarrow 2$ scattering is much more
complicated because in this case, if all gluons are massless, only the
completely collinear scattering is kinematically possible, and the
quantum destructive interference effect with multiple scatterings with
surrounding media called the Landau-Pomeranchuk-Migdal (LPM) effect
should be taken into account.  So, apart from small finite angles
allowed by the effective gluon mass, we can postulate
$\bk_2=k_2\hat{\bk}_1$ and $\bk_3=k_3\hat{\bk}_1$ (where $k_i$
represents not four-vector but $k_i=|\bk_i|$ in expressions below) in
the lower processes in \fref{fig:feyn}.  The collision term for the
$1\leftrightarrow 2$ scattering involves two different kinds of
contributions corresponding to two diagrams in \fref{fig:feyn}.  Now,
since the vector directions of $\bk_2$ and $\bk_3$ are fixed, we can
readily take the angle integrations to express the collision term as
\begin{eqnarray}
\fl
&& \biggl(\frac{\partial f(\bk_1)}{\partial t}
  \biggr)_{1\leftrightarrow 2} = \frac{(2\pi)^3}{2|\bk_1|^2\nu}
  \int_0^\infty \rmd k_2 \rmd k_3 \nonumber\\
\fl
&&\qquad \times \Bigl[ \delta(k_1-k_2-k_3)
    \gamma(\bk_1;\bk_2,\bk_3) \Bigl\{f_2 f_3 (1+f_1)
    - f_1 (1+f_2)(1+f_3) \Bigr\} \nonumber\\
\fl
&&\qquad\quad + 2 \delta(k_1+k_2-k_3) \gamma(\bk_3;\bk_1,\bk_2)
    \Bigl\{f_3 (1+f_1) (1+f_2)
    - f_1 f_2 (1+f_3) \Bigr\} \Bigr] \;.
\label{eq:onetwo}
\end{eqnarray}
The scattering rate $\gamma(\bk_i;\bk_j,\bk_k)$ should contain
multiple interactions with media and should reproduce the
leading-order LPM effect.  The explicit form is given in
\cite{Arnold:2002zm} in a form of the integral equation.

\begin{figure}
 \begin{center}
 \includegraphics[width=0.55\textwidth]{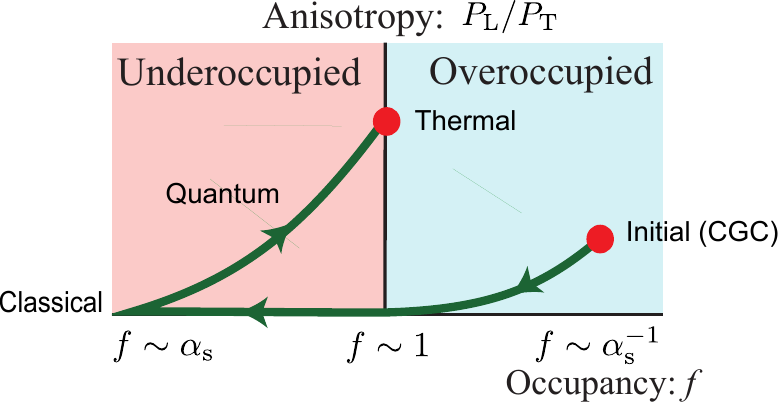}
 \end{center}
 \caption{Schematic paths from the CGC-type initial condition to the
   thermalized state.  Figure is adapted from a talk by Kurkela at
   Quark Matter 2015.  (The original one plots $\PT/\PL$.)}
 \label{fig:kurkela}
\end{figure}

It is not easy to solve these functional equations numerically, and
the state-of-the-art numerical simulation with these equations has
been carried out in \cite{Kurkela:2015qoa}.  The central message from
\cite{Kurkela:2015qoa} is summarized in a schematic picture in
\fref{fig:kurkela}, which is adapted from a picture presented by
Kurkela at Quark Matter 2015.  We note that the original figure plots
$\PT/\PL$ and here the vertical axis is changed to $\PL/\PT$ which is
more consistent with what we have discussed so far.

%---   figure   ---%
%---   figure   ---%

According to the scenario in \fref{fig:kurkela}, $\PL/\PT$ initially
decreases due to longitudinal expansion and it would go to the
free-streaming limit unless the scattering effects are taken into
account.  It is the quantum effect incorporated in the collision terms
that derives the system back to non-zero $\PL/\PT$ and eventually the
system approaches a thermal state.  It is actually a vital question
what lets the system resist against the free-streaming limit, and
there is not a consensus in the heavy-ion physics community yet,
though the quantum fluctuations certainly play a key role.

Another profitable treatment of the collision term is to take the
small angle limit assuming that massless gluon exchange is most
enhanced there.  Specifically, $t$- and $u$-channel terms in the
$2\leftrightarrow 2$ scattering of \eref{eq:Mtwotwo} become dominant,
and in this limiting situation the collision term takes an amazingly
simple form \cite{Blaizot:2013lga}, which has been developed in a
context to address the question of the gluonic BEC formation
speculated in \cite{Blaizot:2011xf}.  We will discuss this possibility
of the BEC formation in details in \sref{sec:thermal} and we here take
a quick look at the concrete form of the collision term.  In
\cite{Blaizot:2013lga} a variation of the QCD Boltzmann equation that
behaves like a Fokker-Planck equation has been proposed with the
collision term,
\begin{equation}
\hspace*{-1em}
  \biggl(\frac{\partial f(\bk)}{\partial t}
   \biggr)_{2\leftrightarrow 2}^{\theta\approx 0}
   = 2\pi^2 \alphas^2 \xi \bnabla_k \cdot\biggl\{ I_a\bnabla_k f(\bk,t)
     +\hat{\bk} I_b f(\bk,t)\bigl[ 1+f(\bk,t) \bigr]\biggr\} \;,
\label{eq:smallangle}
\end{equation}
where $\bnabla_k\equiv \partial/\partial\bk$.  The overall factor
$\xi$ is divergent and requires the UV and the IR cutoffs;
$\xi\equiv (18/\pi)\int_{q_{\rm min}}^{q_{\rm max}}\rmd q/q$ with
$q_{\rm max} \sim T$ and $q_{\rm min} \sim \mD (\sim gT)$.  Here,
$I_a\equiv \int_{\bk} f(\bk)[1+f(\bk)]$ and
$I_b\equiv \int_{\bk} 2f(\bk)/|\bk| \propto\mD$.  Obviously there is
no resummation corresponding to the LPM effect since
\eref{eq:smallangle} corresponds to only $2\leftrightarrow 2$
scattering.  Let us see some interesting properties of
\eref{eq:smallangle}.  First, we can easily check that the
Bose-Einstein distribution function $f_{\rm eq}(\bk)$ leads to a
relation, $I_a=T I_b$, so that \eref{eq:smallangle} vanishes for
$f_{\rm eq}(\bk)$ in equilibrium as it should.  Second, the particle
number obtained by the phase-space integral of $f(\bk,\bx)$ is
conserved manifestly due to the fact that the right-hand side in
\eref{eq:smallangle} is a total derivative.  Therefore, even though
\eref{eq:smallangle} looks very simple, it maintains the essence of
the genuine $2\leftrightarrow 2$ collision term in \eref{eq:twotwo}.
In the original discussions in \cite{Blaizot:2013lga} inelastic
processes are turned off (apart from some qualitative remarks) and the
gluon number is assumed to be a conserved quantity, which inevitably
results in overpopulated gluons and an associated BEC formation.  The
effect of inelastic scattering has been later investigated and in a
recent work \cite{Blaizot:2016bsg} the splitting kernel is simplified
into a form similar to the one in the RTA in \eref{eq:RTA}.  Regarding
the BEC scenario and possible scaling solutions, more discussions will
follow in \sref{sec:thermal}.

The fate of $\PL/\PT$ has been also extensively investigated not only
in the dilute regime but also in the dense regime in terms of
classical field simulations.  In fact, when the occupation number
becomes as large as $f\sim 1/\lambda$ in the $\lambda\phi^4$ theory or
$f\sim 1/\alphas$ in QCD, it is no longer legitimate to utilize the
perturbation theory even for small coupling.  In the dilute regime,
usually, many-body scatterings like $m\leftrightarrow n$ processes are
higher order with respect to the coupling constant.  For example, a
$3\leftrightarrow 3$ scattering in the $\lambda\phi^4$ is of order
$\sim \lambda^4$ at the tree level, and the collision term involves
five distribution functions, leading to the order of $\sim 1/\lambda$
in the saturated regime, which is of the same order as
$2\leftrightarrow 2$ in \eref{eq:twotwo}.  In this saturated regime we
need to use a non-perturbative method such as the semi-classical
approximation.

%%%   Dense regime --- classical statistical simulation   %%%
\subsubsection{Dense regime --- classical statistical simulation}
\label{sec:dense}
The glasma instability was found in the purely classical simulation
with $\dcalA^\mu(\tau,\nu\neq 0)$, and in the first simulation
\cite{Romatschke:2005pm} the initial value of
$\dcalA^\mu(\tau=\tau_0,\nu)$ was treated as white noise proportional
to some seed strength $\Delta$.  For more quantitative studies,
however, we should figure out what the realistic spectrum of
$\Delta(\tau_0,\nu)$ is.  The first attempt along these lines is found
in \cite{Fukushima:2006ax} based on an analogy to the harmonic
oscillator problem in quantum mechanics.

It would give us some intuition if we consider the semi-classical
approximation first not in quantum field theory but in quantum
mechanics.  Let us explain the idea with a simple example, which can
be easily generalized later to quantum field theory problems.  For a
given density matrix $\hat{\rho}(t)$, the Wigner function is defined
as
\begin{equation}
 W(\bx,\bp;t) \equiv \frac{1}{(2\pi\hbar)^3}\int \rmd \delta\bx\,
  \langle\bx-\half\delta\bx| \hat{\rho}(t)
  |\bx+\half\delta\bx\rangle\, \rme^{-\rmi \bp\cdot\delta\bx/\hbar}\;.
\label{eq:Wigner}
\end{equation}
If $\hat{\rho}$ is a pure state of the one-dimensional harmonic
oscillator ground state $|\psi_0\rangle$, and then the wave-function
in the $x$ representation is a Gaussian;
$\psi_0(x)\sim \exp(-x^2/2b^2)$.  It is then straightforward to
confirm $W(x,p)\sim \exp(-x^2/b^2 - b^2 p^2)$.  Thus, roughly
speaking, the Wigner function embodies a probability distribution for
classical conjugate variables in a way consistent with the uncertainty
principle.  It should be mentioned, however, that the Wigner function
as defined in \eref{eq:Wigner} could take a negative value (usually
when some quantum entanglement is involved), and so a naive
interpretation as a probability distribution needs caution.  It can be
proved that a smeared Winger function (called the Husimi function) is
always non-negative, which is a quite useful property to make a
correspondence between the classical fields and the classical particle
distributions \cite{Kunihiro:2008gv}.

From the von~Neumann equation,
$\rmi\hbar (\partial\hat{\rho}/\partial t) = [\hat{H},\hat{\rho}]$,
the time evolution of the Wigner function is determined with the Moyal
product as
\begin{equation}
 \frac{\partial W(\bx,\bp;t)}{\partial t} =
  H(\bp,\bx)\,\frac{2}{\rmi\hbar}\,\sin\Bigl[ \frac{\rmi\hbar}{2}
  \bigl(\overleftarrow{\partial_{\bp}}\,\overrightarrow{\partial_{\bx}}
  -\overleftarrow{\partial_{\bx}}\,\overrightarrow{\partial_{\bp}}
  \bigr)\Bigr]\,W(\bx,\bp;t) \;,
\end{equation}
where the classical Hamiltonian appearing above reads:
\begin{equation}
 H(\bx,\bp) = \int\rmd\delta\bx\, \langle\bx-\half\delta\bx|
  \hat{H}|\bx+\half\delta\bx\rangle\, \rme^{-\rmi\bp\cdot\delta\bx/\hbar}\;.
\end{equation}
We can then expand the above equation of motion in terms of $\hbar$ to
find that, at the leading order, the time evolution is described by
the classical equation of motion with the Poisson brackets:
\begin{equation}
 \frac{W(\bx,\bp;t)}{\partial t}
  = \{ H(\bp,\bx),\, W(\bx,\bp,t)\}_{\rm P} + O(\hbar^2)\;,
\end{equation}
and there is no term of $O(\hbar)$.  Therefore, at least at the
$O(\hbar)$ accuracy, the initial Wigner function has all the quantum
effects and the classical equations of motion remain intact.  This
observation is the theoretical foundation of the classical statistical
simulation.

Historically, the classical statistical simulation has been developed
in a wider context than the heavy-ion collision physics.  A successful
example in the thermalization problem in the Early Universe is found
in \cite{Micha:2002ey} where turbulent behavior with self-similar
dynamics has been observed in semi-classical $\lambda\phi^4$ theory
(see also \cite{Micha:2004bv} for a more comprehensive report).  On
a more academic level, a scalar theory with high initial occupancy was
considered in \cite{Aarts:2001yn} to quantify classical aspects of
real-time quantum dynamics, and the classical statistical formulation
for non-Abelian gauge theories was given in \cite{Berges:2007re}.
There are fruitful outputs from the classical statistical simulation,
especially many insightful indications about the fate of the
isotropization (discussed more in \sref{sec:isotropization}) and the
weak wave turbulence in the pure Yang-Mills theory (discussed more in
\sref{sec:thermal}).  Interested readers are guided to the most recent
review \cite{Berges:2015kfa} on the classical statistical simulation.

For the rest of this subsubsection, let us explain how the initial
Wigner function should be given for a special geometry in the
heavy-ion collisions with longitudinal expansion.  This problem was
carefully resolved in \cite{Dusling:2011rz} and further investigated
for a scalar theory in \cite{Dusling:2012ig} and for gauge field
theories in \cite{Epelbaum:2013waa}.  Here, let us take a close look
at the derivation of fluctuation spectrum in an expanding scalar
theory defined with a Lagrangian density,
$\calL(\phi) = \frac{1}{2}(\partial^\mu\phi)^2 - V(\phi)$.  To mimic
the glasma background, let us decompose a scalar field $\phi$ at small
$\tau$ into an $\eta$-independent background $\varphi$ (which can be
regarded as $\tau$-independent for $\tau\approx 0$) and
$\eta$-dependent quantum fluctuations as
\begin{equation}
\fl
\phi(\bxt,\eta,\tau\approx 0) = \varphi(\bxt) + \frac{\sqrt{\pi}}{2}
  \int\frac{\rmd\nu}{2\pi}\rmd \mu_K\; \rme^{\pi\nu/2}\,
  c_{\nu K}\,\rme^{\rmi\nu\eta} \chi_K(\bxt)
  H_{\rmi\nu}^{(2)}(\lambda_K\tau) + \mbox{c.c.}\;,
\label{eq:spect}
\end{equation}
and the question is the probability distribution for the weight
$c_{\nu K}$ of each mode.  Here, $H_{\rmi\nu}^{(2)}(x)$ is the Hankel
function and $\chi_K(\bxt)$ represents the orthogonal basis function
on top of the background $\varphi$, which is determined by the
eigenvalue equation,
\begin{equation}
  \bigl[ -\bnabla_\perp^2 + V''(\varphi) \bigr] \chi_K(\bxt)
  = \lambda_K^2\, \chi_K(\bxt)\;,
\end{equation}
and $K$ is the quantum number to label different eigen-vectors.  If
the background potential $V''(\varphi)$ is spatially uniform, $K$ is
nothing but a spatial momentum $\bk$ and $\chi_K(\bxt)$ is a plane
wave.  The choices of the eigen-function and the measure $\rmd\mu_K$
are not independent;  a choice proposed in \cite{Dusling:2012ig} is,
using
\begin{equation}
  \delta_{K K'} \equiv \int\rmd^2\bxt\, \chi_K^\ast(\bxt)\,
  \chi_{K'}(\bxt) \;,
\end{equation}
the measure is normalized to satisfy,
\begin{equation}
  \int\rmd\mu_K\, \delta_{K K'} = 1\;.
\end{equation}
Then, the spectrum of initial quantum fluctuations is characterized in
the following form:
\begin{equation}
  \langle c_{\nu K}\, c_{\mu K'}\rangle = 0\;,\qquad
  \langle c_{\nu K}\, c_{\mu K'}^\ast \rangle = \pi \delta(\nu-\mu)
  \delta_{KK'}\;.
\end{equation}
It might be a bit puzzling why \eref{eq:spect} involves
$H_{\rmi\nu}^{(2)}(\lambda_K\tau)$ even though we are interested only
in the initial spectrum at $\tau=0^+$.  The reason is that there are
coordinate singularities at $\tau=0^+$ and for practical simulations
we need to start the numerical simulation with some initial condition
at small but finite $\tau$.

It is non-trivial how to define the occupation number from the
classical fields.  In other words, because the occupation number is an
expectation value of the number operator in terms of the annihilation
and the creation operators, what we need is the representation of the
annihilation and the creation operators using the classical fields.
This can be done with the projection to free particle basis,
i.e.\ (see \cite{Fukushima:2014sia} for a related argument in the
context of the Schwinger mechanism)
\begin{equation}
\fl
  f_{\nu\bkt}(\tau) = -\frac{1}{2}
  + \frac{\pi\tau^2\rme^{\pi\nu}}{4S_\perp L_\eta} \biggl\langle
  \biggl| \int\rmd^2\bxt \rmd\eta\,\rme^{-\rmi\nu\eta-\rmi\bkt\cdot\bxt}
  \, H_{\rmi\nu}^{(2)\ast}(k_\perp\tau)
  \overset\leftrightarrow\partial \phi(\tau,\eta,\bxt)\biggr|^2
  \biggr\rangle\;.
\end{equation}
This formulation of the classical statistical simulation with correct
quantum spectrum should reproduce at least the one-loop order
results.  The advantage lies in the stability for long time
simulations, while serious shortcomings are found in the UV sector
when applied for quantum field theories.  First of all, the zero-point
oscillation energy appears and it should be gotten rid of by some
subtraction procedures.  In ordinary quantum field theory the
zero-point oscillation energy is just an offset in energy and safely
discarded.  The situation gets highly complicated as soon as
inhomogeneous background fields are involved especially in expanding
geometries.  One prescription would be to take a finite difference
between numerical results with and without the background fields, as
was implemented in \cite{Gelis:2013rba}.  Secondly, the approximation
in the classical statistical simulation may ruin the renormalizability
of theory, which was shown perturbatively in \cite{Epelbaum:2014yja}.

To have a deeper insight into field theoretical problems inherent in
the classical statistical simulation, it would be very useful to
understand how the classical description can have a connection to the
kinetic equation when the occupation number gets large.  This question
was formulated for $\lambda\phi^4$ theory in \cite{Mueller:2002gd} and
some subtleties in the derivation have been clarified in
\cite{Jeon:2004dh}.  Here, let us take a quick look over the arguments
in \cite{Mueller:2002gd}.  To make the question well-defined, we
should work in a semi-saturated regime where $1\ll f\ll 1/\lambda$ is
assumed.

The key elements are the real-time propagators $G_{ij}(x,y)$ in the
so-called $ra$ basis.
In non-equilibrium case the translational
invariance could be violated and so, generally speaking, $G_{ij}(x,y)$
is a function of not only $\delta x=x-y$ but also $X=(x+y)/2$.
Assuming that the $X$ dependence is slow, replacing $X$ with $x$, the
Fourier transformed propagators with respect to $\delta x$ can be
expressed as
\begin{eqnarray}
\fl\qquad
  G_{aa}(k,x) &=& 0\;,\quad
  G_{ra}(k,x) = \frac{\rmi}{k^2\!-\!m^2\!+\!\rmi \epsilon k_0} \;,\quad
  G_{ar}(k,x) = \frac{\rmi}{k^2\!-\!m^2\!-\!\rmi \epsilon k_0} \;,\\
\fl\qquad
  G_{rr}(k,x) &=& \biggl[\frac{1}{2}+f(\bk,\bx,t)\biggr] \bigl(
    G_{ra}-G_{ar}\bigr)
  = \biggl[\frac{1}{2}+f(\bk,\bx,t)\biggr]\,2\pi\delta(k^2-m^2)\;.
\end{eqnarray}
These propagators should satisfy the Dyson equation:
\begin{equation}
  2\rmi k^\mu\partial_\mu G_{rr}(k,x) = G_{rr}(\Sigma_{ar}-\Sigma_{ra})
  + \Sigma_{aa}(G_{ar}-G_{ra}) \;.
\end{equation}
Because $f$ is large now, $G_{rr}(p,x)$ is dominant, and then the
Dyson equation leads to a kinetic equation with the collision term in
this approximation given by
\begin{equation}
  \biggl(\frac{\partial f}{\partial t}
    \biggr)_{\rm coll}^{\rm classical}
  = \frac{-\rmi (\Sigma_{ar}-\Sigma_{ra})}{2k_0}
  \Bigl( f+\frac{1}{2}\Bigr) + \frac{\rmi\Sigma_{aa}}{2k_0}\;.
\end{equation}
The self-energies can be computed according to the ordinary Feynman
rule in the $\lambda\phi^4$ theory.  Because all self-energies are
written in terms of $G_{rr}\propto (1/2+f)$, we can understand that
the collision term for the $2\leftrightarrow 2$ scattering is modified
from the conventional form of \eref{eq:twotwo} into
\begin{eqnarray}
\fl
&& \biggl(\frac{\partial f(\bk_1)}{\partial t}
  \biggr)_{2\leftrightarrow 2}^{\rm classical} = \frac{\lambda^2}{4\omega(\bk_1)}
  \int_{\bk_2,\bk_3,\bk_4}
  (2\pi)^4 \delta^{(4)}(k_1+k_2-k_3-k_4)\, \nonumber\\
\fl
&& \quad \times \Bigl\{(f_3+\half)(f_4+\half)(f_1+\half)
 + (f_2+\half)(f_3+\half)(f_4+\half) \nonumber\\
&& \qquad - (f_1+\half)(f_2+\half)(f_3+\half)
 - (f_4+\half)(f_1+\half)(f_2+\half) \Bigr\}\;,
\label{eq:twotwoclass}
\end{eqnarray}
where the cubic terms and the quadratic terms reproduce the correct
ones in \eref{eq:twotwo}, while this above form has extra linear
terms, $\propto f_3+f_4-f_2-f_1$.  Surprisingly, the presence of these
linear terms change the structure of theory in the UV region
drastically \cite{Epelbaum:2014mfa}.  To see this, let us consider the
equilibrium distribution resulting from \eref{eq:twotwoclass}, that
is easily found to be the Rayleigh-Jeans form:
\begin{equation}
  f(\bk) \to \frac{T}{\omega(\bk)-\mu} - \frac{1}{2}\;,
\label{eq:expand_therm}
\end{equation}
which correctly reproduces first two terms from the expansion of the
Bose-Einstein or Planck distribution.  Therefore, this is a valid
description for $f\gg 1$ with $\omega-\mu \ll T$.  For large $|\bk|$,
however, $f(\bk)$ becomes smaller and smaller and eventually the
approximation breaks down.  In the genuine thermal equilibrium,
$f(\bk)$ should have an exponential tail rather than a power-law
decay, which cannot be reproduced in the semi-classical
approximation.

This change to \eref{eq:expand_therm} is the clearest manifestation of
the loss of renormalizability.  In fact, with the distribution
function \eref{eq:expand_therm}, the total particle number and the
energy are both UV divergent (i.e.\ \textit{UV catastrophe}, which is
a well recognized problem in the condensation of classical non-linear
waves \cite{PhysRevLett.95.263901}) and a UV cutoff is necessary even
though the underlying theory was originally renormalizable.  This
implies that the classical statistical simulation should suffer
artificial dependence on a UV cutoff, which moreover affects the
scaling behavior \cite{Epelbaum:2015vxa}, as is the main subject in
\sref{sec:isotropization}.

%%%%%   Other Methods   %%%%%
\subsection{Other Methods}
\label{sec:unconventional}
It would be desirable to invent theoretical methods applicable to both
dilute and dense regimes particularly to investigate the whole
dynamics of an expanding system in the heavy-ion collision.  There is
unfortunately no such universal method so far, but theoretical
attempts are making some progresses, some selected ones of which will
be introduced in this section.

%%%   Kadanoff-Baym equations   %%%
\subsubsection{Kadanoff-Baym equations}
The quantum upgraded version of the equations of motion is the
Dyson-Schwinger equation.  In principle the kinetic equation could be
derived from the Dyson-Schwinger equation.  It is still very difficult
to solve the Dyson-Schwinger equation or similar functional equations
in Minkowskian spacetime (see \cite{Gasenzer:2008zz} for an attempt
based on functional renormalization group equations);  a part of
subtlety comes from a technical difficulty in imposing a UV cutoff to
Minkowskian four-vector.  It would be a better strategy to transform
the functional equation into a more convenient representation such as
the 2PI formalism \cite{Cornwall:1974vz}.  In the context of the
real-time studies the 2PI formalism has been successful for the $1/N$
expansion in $O(N)$ scalar theories as discussed diagrammatically in
\cite{Aarts:2002dj} and numerically with instability in
\cite{Berges:2008wm}.

The 2PI effective action (for a bosonic field) reads:
\begin{equation}
  \Gamma[G] = \frac{\rmi}{2} \Tr_C\bigl[\ln G^{-1}
    + G_0^{-1}G\bigr] + \Gamma_2[G]\;,
\end{equation}
where $G_0$ and $G$ represent the free and the full propagators,
respectively, and $\Gamma_2[G]$ is the contribution from 2PI diagrams
in terms of bare vertices and full propagators.  The trace $\Tr_C$ is
taken along the closed-time path.  The full propagator and the
self-energy are determined functionally from stationary conditions,
\begin{equation}
  \frac{\delta\Gamma[G]}{\delta G} = 0\;,\qquad
  \Pi = 2\rmi\frac{\delta\Gamma_2[G]}{\delta G}\;,
\end{equation}
which yield the Kadanoff-Baym equations.  The propagator and the
self-energy are decomposed into statistical and spectral parts;
$G(x,y)=G_F(x,y)-\frac{\rmi}{2}{\rm sign}_C(x^0-y^0)G_\rho(x,y)$ and
$\Pi(x,y)=-\rmi\delta_C(x-y)\Pi^{\rm (local)}(x)+\Pi_F(x,y)
-\frac{\rmi}{2}{\rm sign}_C(x^0-y^0)\Pi_\rho(x,y)$, where
${\rm sign}_C(x^0-y^0)\equiv\Theta_C(x^0-y^0)-\Theta_C(y^0-x^0)$.
The Kadanoff-Baym equations
read \cite{Berges:2002wt}:
\begin{eqnarray}
\fl
\bigl[ \partial^2 + M^2(x)\bigr]G_F(x,y)
 = \int_0^{y^0} \rmd^4 z\,\Pi_F(x,z)G_\rho(z,y)
  - \int_0^{x^0} \rmd^4 z\,\Pi_\rho(x,z) G_F(z,y)\;.\\
\fl
\bigl[ \partial^2 + M^2(x)\bigr]G_\rho(x,y)
  = -\int_{y^0}^{x^0}\rmd^4 z\,\Pi_\rho(x,z) G_\rho(z,y)
\end{eqnarray}
with $M^2(x)\equiv m^2+\Pi^{\rm (local)}(x)$.  The Wigner transformed
propagators are defined as
$\tilde{G}_F(X,k)=\int\rmd^4\delta x\,\rme^{\rmi k\delta x}
G_F(X+\frac{\delta x}{2},X-\frac{\delta x}{2})$ and
$\tilde{G}_\rho(X,k)=-\rmi\int\rmd^4\delta x\,\rme^{\rmi k\delta x}
G_\rho(X+\frac{\delta x}{2},X-\frac{\delta x}{2})$.  The gradient
expansion leads to a Boltzmann-type kinetic equation for the
distribution function $f(X,k)$ where it is defined from
$\tilde{G}_F=\tilde{G}_\rho(f+\frac{1}{2})$ and the quasi-particle
approximation, $\tilde{G}_\rho=\pi\bigl[\delta(k^0-\varepsilon(X,\bk))
  -\delta(k^0+\varepsilon(X,\bk))\bigr]/\varepsilon(X,\bk)$ with
$\varepsilon(X,\bk)=\sqrt{\bk^2+M^2(x)}$, is used.

A qualitative difference between the Boltzmann equation and the
Kadanoff-Baym equation appears from the quasi-particle approximation,
without which the collision phase space opens for
$0\leftrightarrow 4$, $1\leftrightarrow 3$, $2\leftrightarrow 2$
off-shell processes in the $\lambda\phi^4$ theory.  The numerical
simulation of the QCD Kadanoff-Baym equation is still an ambitious
challenge.  To simplify the treatment of the spectral function,
$\tilde{G}_\rho$, a Lorentzian Ansatz was introduced in some
phenomenological approaches (see \cite{Cassing:2008nn} for a review),
and the more full self-consistent treatment was carried out in
\cite{Hatta:2011ky} with the aim to make a unified formalism with the
CGC background.  This direction of research should deserve more
studies with computer resource invested in the future.

%%%   Stochastic quantization   %%%
\subsubsection{Stochastic quantization}
The Monte-Carlo integration is useless when the sign problem is
severe, and this is why the direct QCD simulation is so difficult in
Minkowskian spacetime.  Then, one idea to overcome the sign problem is
to quantize a field theory in a different way, using a Langevin
equation, which is conceivable because quantum effects are
fluctuations around the classical paths.  One of the oldest attempts
along these lines is the derivation of the Schr\"{o}dinger equation
from a Brownian motion by Nelson \cite{Nelson:1966sp}.  It is known
that classical noises are inadequate for correct quantization, but it
is possible to reformulate the quantization procedure by adding a
fictitious time or a quantum axis.  This method is thus an example of
the so-called \textit{holographic principle} that states an
equivalence between $D$-dimensional quantum theory and
$(D+1)$-dimensional classical theory.

A complete review is available in \cite{Damgaard:1987rr}; here, we
simply sketch the idea.  For a simple scalar theory, the Langevin
equation to describe the evolution with the fictitious time $\theta$
is written down as
\begin{equation}
  \partial_\theta\phi(x,\theta) = \rmi\frac{\delta S}{\delta \phi(x)}
  \biggr|_{\phi(x)\to\phi(x,\theta)} + \eta(x,\theta)\;,
\label{eq:Langevin}
\end{equation}
where $S$ is an action to define the theory and $\eta(x,\theta)$ is a
stochastic noise satisfying
$\langle\eta(x,\theta)\eta(x',\theta')\rangle_\eta=2\delta^{(4)}(x-x')
\delta(\theta-\theta')$.  It is claimed that the quantum expectation
value of an operator $\calO[\phi(x)]$ is given by
\begin{equation}
  \langle\calO[\phi(x)]\rangle = \lim_{\theta\to\infty}
  \langle\calO[\phi(x,\theta)]\rangle_\eta\;.
\end{equation}
Using the stochastic diagrams, we can map the above procedures
faithfully to the conventional Feynman diagrams.  Therefore, the
perturbative equivalence has no doubt based on diagrammatic
considerations, while the non-perturbative simulations could violate
this perturbative equivalence.

Because the Langevin equation in \eref{eq:Langevin} is complex with an
imaginary unit in front of the drift term, this quantization procedure
is nowadays called the complex Langevin method.  The first successful
report on the real-time quantization is \cite{Berges:2005yt}, in which
the boundary condition was not correctly implemented, and a more
refined simulation was performed in \cite{Berges:2007nr}.  The method
seemed to be promising apart from a stability problem in a long-time
simulation.  Later, the real-time complex Langevin method was
revisited in \cite{Anzaki:2014hba} and it was found that the numerical
simulation has a general tendency to fall into a wrong answer.

New insights to the complex Langevin method have emerged from careful
analyses in comparison to the Lefschetz thimble method that looks
similar to the complex Langevin method but has a firm mathematical
foundation.  The relation between these two methods has been
understood numerically \cite{Aarts:2014nxa} and analytically
\cite{Fukushima:2015qza}, which was useful to clarify the origin of
the convergence problem in the complex Langevin method.  In general,
when the Stokes phenomenon occurs in complexified theories, the
convergence becomes subtle, which can be understood from phase factors
of distinct Lefschetz thimbles \cite{Hayata:2015lzj}.  Usually the
Stokes phenomenon corresponds to a phase transition in equilibrium
environments and so the complex Langevin method works poorly only when
the system approaches a phase transition \cite{Fujii:2015bua}.  In
Minkowskian spacetime the situation is much worse and the onset of the
Stokes phenomenon is found around the on-shell conditions, and the
validity region is tightly limited.  Without some breakthrough, it is
unlikely that the complex Langevin method or the Lefschetz thimble
method can capture the correct real-time dynamics of interested
physics problems.  An important lesson that we can learn is that some
unexpected complication may appear and a different prescription to
quantize a theory may change non-perturbative contents of the theory,
which could be understood from a well-known mathematical fact that
many inequivalent functions can happen to have identical asymptotic
series.

%%%   Gauge/gravity correspondence   %%%
\subsubsection{Gauge/gravity correspondence}
The most widely recognized example of the holographic principle is the
correspondence between the gauge theory and the gravity theory, i.e.,
the $\mathcal{N}=4$ supersymmetric Yang-Mills theory in the
large-$\Nc$ limit and the classical solution (anti-de~Sitter; AdS$_5$
metric) of the super-gravity theory.  In the heavy-ion community this
method has become very popular since the successful calculation of the
shear viscosity \cite{Policastro:2001yc}.

The key relation of the correspondence is summarized in a form of the
GKP-Witten relation;
\begin{equation}
  \biggl\langle \exp\biggl[\rmi \int\rmd^4 x\,\phi_0(x)\calO(x)\biggr]
  \biggr\rangle = \rme^{\rmi S_{\tiny \mbox{gravity}}[\phi_0(x)]}\;,
\end{equation}
where the left-hand side is the expectation value in gauge field
theory and the right-hand side is the on-shell action in the classical
gravity theory with the boundary condition $\phi\to\phi_0$ at the
boundary.  Apart from decoupled and irrelevant coordinates in $S^5$,
the fifth coordinate $z$ in addition to Minkowskian $t$ and $\bx$
refers to the quantum axis, which together span AdS$_5$ space.  In the
gravity side the theory is described by the equations of motion in a
bulk from $z=\infty$ (UV) toward $z=0$ (IR) and the gauge theory
resides in a boundary at $z=0$;  in this sense, the gauge/gravity
correspondence could be regarded as the bulk/boundary correspondence
or the UV/IR correspondence.

The very first application of the gauge/gravity correspondence to
investigate the early-time dynamics in the heavy-ion collision is
a work by Janik and Peschanski in \cite{Janik:2005zt}, which was
re-derived also in \cite{Kovchegov:2007pq}.  For a pedagogical
introduction, a review by Peschanski \cite{Peschanski:2008xn} should
be quite readable for ``users'' of this string-inspired technique.

Like the lattice-QCD simulation, the gauge/gravity correspondence is a
powerful method to compute an expectation value of gauge invariant
operator, and for the early-time dynamics in the heavy-ion collision,
the most informative observable is the energy-momentum tensor
$T_{\mu\nu}$.  What we should do first is to obtain the solution of
the 5-dimensional gravity equations as
$\rmd s^2=[g_{\mu\nu}(z)\rmd x^\mu \rmd x^\nu + \rmd z^2]/z^2$
in the Fefferman-Graham form (choosing appropriate coordinates).
Then, the energy-momentum tensor is inferred from the relation,
\begin{equation}
  g_{\mu\nu}(z\approx 0,x) = \eta_{\mu\nu}
  + \frac{2\pi^2}{\Nc^2}\langle T_{\mu\nu}(x)\rangle\,z^4 + \cdots \;.
\end{equation}
In \cite{Janik:2005zt,Kovchegov:2007pq} a black-hole solution has been
discovered that corresponds to the one-dimensional expansion of
hydrodynamics (i.e.\ the Bjorken solution).  The most interesting
finding is that the time dependence of the energy density
$\varepsilon(\tau)$ is a constant initially and turns to $\tau^{-4/3}$
later; this latter scaling recovers the fully isotropized
hydrodynamical one in \eref{eq:Bjorken_scaling}.  The transitional
change from a constant to $\tau^{-4/3}$ behavior should be identified
as the isotropization point, which yields an isotropization time scale
as
\begin{equation}
   \tau_{\rm iso} =
   \biggl(\frac{3\Nc^2}{2\pi^2 e_0}\biggr)^{3/8}\;,
\end{equation}
where $e_0$ is defined through the initial energy density
$\varepsilon_0=e_0\tau_0^{4/3}$ at $\tau=\tau_0$.  If we consider
$\varepsilon_0\sim 15\GeV/\fm^3$ at $\tau_0=0.6\fm/c$ at RHIC energy,
we can have an estimate at strong coupling as $\tau_{\rm iso}\simeq
0.3\fm/c$, which might be an account for the fast isotropization.

Instead of postulating a black hole solution corresponding to
dynamical QGP, the heavy-ion collision itself could be emulated by a
shock-wave collision in the gravity side, which looks like a CGC-like
problem to solve the classical equations of motion with two colliding
sources (and like the CGC setup the one shock-wave problem is
analytical solvable; see a pedagogical review \cite{Janik:2010we} and
references therein).  The pioneering numerical work to simulate the
horizon (and QGP in a gauge dual side) formation is found in
\cite{Chesler:2008hg}, in which $\PT$ and $\PL$ have been calculated
as functions of time.  Interestingly, right after the collision, $\PT$
goes positively and $\PL$ goes negatively, in a way similar to the CGC
simulation.  This implies that a picture of extending color flux tubes
in the CGC initial condition should be the right physics description
for the very early dynamics.  However, in the gauge/gravity numerical
simulation, it has been observed that $\PL/\PT\to 1$ as quickly as
$\tau_{\rm iso}\sim 0.7/T\simeq 0.5\fm/c$ for the initial temperature
$T\sim 0.35\GeV$.  Later, in \cite{Chesler:2010bi}, by means of the
holographic numerical solutions, the validity of (first-order) viscous
hydrodynamics has been tested in a region where $\PL/\PT<1$, which was
such an important test that the way of thinking in the heavy-ion
community was changed.  Before \cite{Chesler:2010bi}, there were many
studies on the isotropization, and sometimes it was not clearly
distinguished from the hydrodynamization.  Now, we know that the
viscous hydrodynamics can work well even when strong anisotropy still
remains.  It might sound a bit puzzling that the viscous hydrodynamics
is required for the system described by a gauge/gravity dual in which
the shear viscosity is as small as the unitarity limit and the bulk
viscosity is vanishing.  We will come back to this question in
\sref{sec:hydrodynamics}.

%%%%%%%%%%   More on Isotropization   %%%%%%%%%%
\section{More on Isotropization}
\label{sec:isotropization}

This section is devoted to a status summary of the scaling solution
and its classification that includes a possibility going to the
free-streaming limit.  It is still under dispute what microscopic
dynamics can sustain the system staying away from the vanishing
longitudinal pressure.

%%%%%   Scaling Properties   %%%%%
\subsection{Scaling Properties}
To sort various scenarios out, it is quite useful to introduce a
scaling form of the solution for the gluon distribution as a function
of time.  The self-similarity that has been confirmed in many
classical statistical simulations implies the following scaling
properties for the gluon distribution:
\begin{equation}
  f(\bk,\tau) = (\Qs\tau)^\alpha f_{\rm S}\bigl(
  (\Qs\tau)^\beta k_\perp, (\Qs\tau)^\gamma k_z \bigr)\;,
\label{eq:scaling}
\end{equation}
which was systematically studied in \cite{Berges:2013fga} (which
nicely reviews all technical details including lattice
discretization).  The exponents $\alpha$, $\beta$, and $\gamma$
characterize the non-equilibrium dynamical evolution, and these are
reminiscent of the critical exponents in the vicinity of IR fixed
points on the renormalization flow.  This is a new form of the
\textit{universality} out of equilibrium, and unlike the static
situation, there is no simple classification of the universality class
only according to the dimensionality and the global symmetry.

%---   table   ---%
\begin{table}
  \begin{center}
  \begin{tabular}{c|c|c|c}
    Authors & $\alpha$ & $\beta$ & $\gamma$ \\
    \hline\hline
    BMSS \cite{Baier:2000sb} & $-2/3$ & $0$ & $1/3$ \\
    B \cite{Bodeker:2005nv} & $-3/4$              & $0$ & $1/4$ \\
    BGLMV \cite{Blaizot:2011xf} & $-(3-\delta_s)/7$ &
           $(1+2\delta_s)/7$ & $(1+2\delta_s)/7$ \\
    KM \cite{Kurkela:2011ub} & $-7/8$ & $0$ & $1/8$
  \end{tabular}
  \end{center}
\caption{Different exponents according to different scenarios with
  one-dimensional expansion.  If complete thermalization with isotropy
  is achieved, $\alpha=0$, $\beta=\gamma=1/3$ should be realized.}
\label{tab:exponents}
\end{table}
%---   table   ---%

In the case with one-dimensional expansion, a typical value of $k_z$
should be decreased as $\tau$ elapses, so that $\gamma$ is supposed to
be positive.  In fact, in the free-streaming limit, $\gamma=1$ is
expected.  Quantitative values of $\alpha$, $\beta$, and $\gamma$
strongly depend on the interactions or the collision terms in the
Boltzmann equation.  \Tref{tab:exponents} is a list of exponents with
one-dimensional expansion as discussed in \cite{Berges:2013fga}.
Here, we will not see all the derivations, but focus on the value of
BMSS which refers to the bottom-up thermalization scenario in
\cite{Baier:2000sb}.

As explained in \sref{sec:bottomup} $\Delta k_\perp^2$ was estimated
by $\hat{q}\sim \alphas^2 T^3$ in the bottom-up thermalization.  To
identify the scaling exponent, we must know how $\hat{q}$ should
parametrically depend on the distribution function; that is,
$\hat{q}\sim \int\rmd^2 k_\perp\, k_\perp^2\,\rmd \Gamma/\rmd^2 k_\perp$
where $\rmd\Gamma/\rmd^2 k_\perp$ is the scattering rate
that scales as $\sim \alphas^2 \int\rmd k_z\,f^2$, and eventually we
have $\hat{q}\sim \alphas^2 \int_{\bk} f^2 \sim
(\Qs\tau)^{2\alpha-2\beta-\gamma}$.  Because
$\hat{q}\sim \Delta k_\perp^2/\rmd\tau$, it is conceivable to
postulate the collision term parametrically scaling as
$(\partial f/\partial \tau)_{\rm coll}\sim
(\Delta k_\perp^2/\rmd\tau)\cdot (\partial f/\Delta k^2)
\sim \hat{q}\partial_{p_z}^2 f \sim (\Qs\tau)^{3\alpha-2\beta+\gamma}$,
where the small angle approximation was used to pick only
$\partial_{p_z}$ up from $\nabla_p$ [see also \eref{eq:smallangle}].
From the Boltzmann equation \eref{eq:Boltzmann_Bj}, we can immediately
deduce $\rmd f/\rmd\tau\sim (\Qs\tau)^{\alpha-1}
\sim (\Qs\tau)^{3\alpha-2\beta+\gamma}$, leading to
\begin{equation}
  2\alpha - 2\beta + \gamma = -1\;.
\label{eq:collision_scale}
\end{equation}
As long as the elastic collision is dominant, which is the case for
the scattering processes with $k_{\rm br}\sim \Qs$, the gluon number
is approximately a conserved quantity.  This gives, in the
one-dimensional expanding geometry,
$\mbox{(const)}\sim (\Qs\tau)\int_{\bk} f \sim
(\Qs\tau)^{\alpha-2\beta-\gamma+1}$.  In the same way, the energy
conservation gives another scaling relation.  With an energy quanta
approximated as $\omega(\bk)\sim k_\perp$, which is true for
$\gamma>\beta$ in late time, it is straightforward to see that the
energy conservation and the momentum conservation can be
simultaneously satisfied only when $k_\perp \sim \mbox{(const)}$,
i.e., $\beta=0$.  Therefore, we have two more conditions as
\begin{equation}
  \alpha - 2\beta - \gamma = -1\;,\qquad \beta = 0\;.
\label{eq:conserv}
\end{equation}
Here, we should note that the number conservation is a robust argument
as long as elastic scatterings are dominant, while the energy
conservation is not.  An immediate counter example is the full
thermalized system for which $\alpha=0$, $\beta=\gamma=1/3$ should be
expected, which seems to violate the energy conservation.  In fact,
the energy is lost by the expansion with non-zero longitudinal
pressure, and thus, the above scaling arguments implicitly assume a
situation close to the free-streaming limit.  With these cautions in
mind, we can solve these scaling relations to determine the exponents
uniquely as $\alpha=-2/3$ and $\gamma=1/3$, and this is how BMSS
values in \tref{tab:exponents} are obtained.

It is interesting to point out that BMSS, B, and KM satisfy
\eref{eq:conserv}, which means that the total particle number and the
energy are strictly conserved and the difference in the exponents is
attributed to the concrete form of the collision terms; namely,
\eref{eq:collision_scale} may be changed by various scenarios.
Indeed, if the collision term has another scaling;
$(\partial f/\partial\tau)_{\rm coll}\sim (\Qs\tau)^\mu$, then
\eref{eq:collision_scale} should be replaced with $\mu-\alpha=-1$.

Now, a question may well arise from the exception in
\tref{tab:exponents}; what is assumed in BGLMV that obviously violates
either particle number or energy conservation.  Because we can confirm
that $\alpha-3\beta-\gamma=-1$ holds apart from $\delta_s$ that
represents the effect of expansion, the energy is conserved in this
scenario, while the particle number conservation is abandoned.
Actually, this scenario accommodates a possibility of the BEC
formation and a finite fraction of particles condenses at the zero
mode.  We discuss this speculative picture in details in
\sref{sec:thermal}.

The classical statistical simulation in the pure Yang-Mills theory
favors the BMSS exponents according to the results in
\cite{Berges:2013fga}.  This idea of the universality classification
based on the scaling properties could open a new theoretical scheme to
tackle non-equilibrium statistical physics in general
\cite{Berges:2014bba} and it would be a challenging problem to
establish a complete list of classification, i.e., a counterpart of
the classification of the dynamical critical phenomena as summarized
in \cite{RevModPhys.49.435}.  For our purpose of the isotropization
problem in the heavy-ion collision, though a deviation from $\gamma=1$
certainly suggests non-trivial physics different from the
free-streaming limit, $\PL/\PT$ goes vanishingly small for large
$\tau$ as long as the scaling \eref{eq:scaling} with $\gamma>\beta$ is
the case.

%%%%%   Classical vs. Quantum Simulations   %%%%%
\subsection{Classical vs.\ Quantum Simulations}
The most relevant quantity of our interest in the heavy-ion collision
is the time-dependence of $\PL/\PT$, and it would make sense to
parametrize it as $\PL/\PT\sim (\Qs\tau)^{-\beta_{\rm eff}}$, or
equivalently,
\begin{equation}
  \beta_{\rm eff} \equiv -\tau \frac{\rmd}{\rmd\tau} \ln(\PL/\PT) \;,
\end{equation}
which is an exponent introduced in \cite{Epelbaum:2015vxa}.  Because
the longitudinal/transverse pressure is to be written as
$P_{\rm L/T}(\tau)=\int_{\bk} (k_{\parallel/\perp}^2/(2)|\bk|) f(\bk,\tau)$,
we see $\beta_{\rm eff}\sim 2(\gamma-\beta)$, which goes to
$\beta_{\rm eff}\to 2$ in the free-streaming limit ($\beta=0$ and
$\gamma=1$) and $\beta_{\rm eff}\to 2/3$ in the classical statistical
simulation or in BMSS ($\beta=0$ and $\gamma=1/3$), and supposedly in
realistic physical systems $\beta_{\rm eff}\to 0$ should be the right
answer.  Thus, we can rephrase the isotropization problem as a puzzle
to explain $\beta_{\rm eff}=0$ that has never been realized in
reliable numerical simulations.

%---   figure   ---%
\begin{figure}
 \begin{center}
 \includegraphics[width=0.5\textwidth]{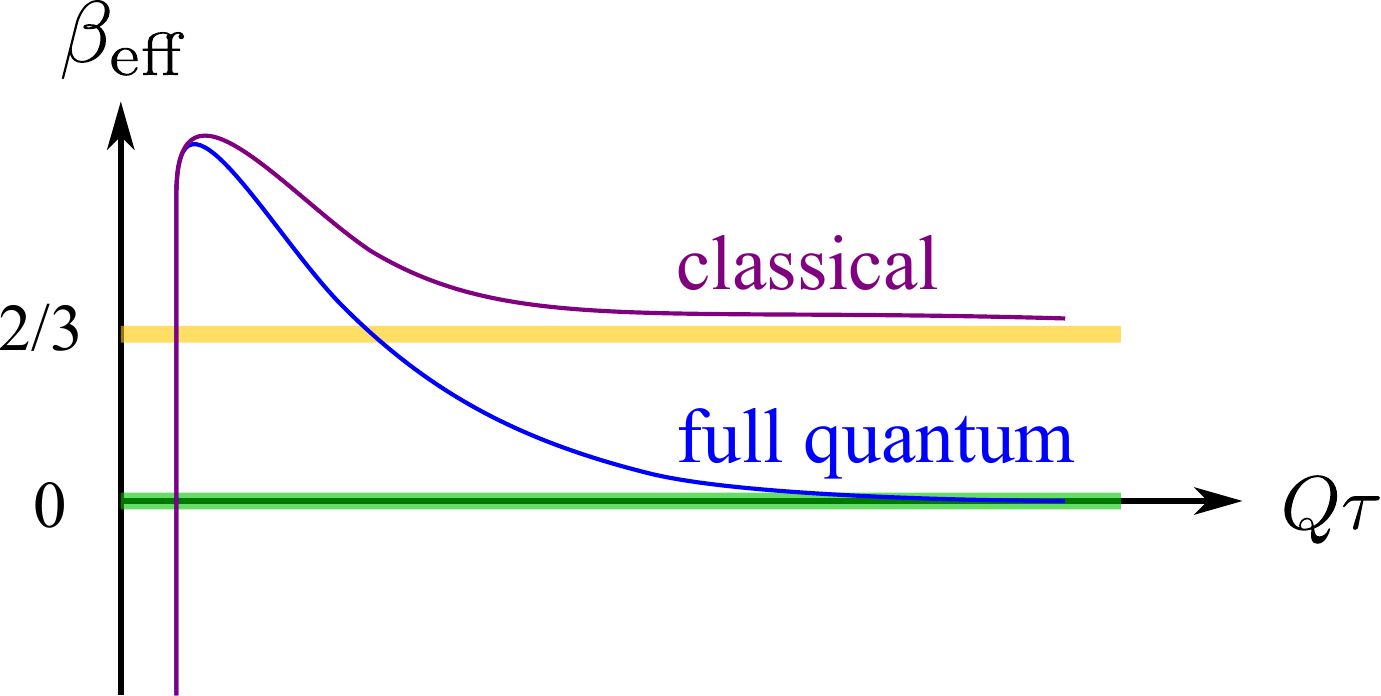}
 \end{center}
 \caption{Schematic picture of the evolution of the exponent
   characterizing the isotropization degree.  Figure sketched based on
   the results in \cite{Epelbaum:2015vxa}.}
  \label{fig:qboltzmann}
\end{figure}
%---   figure   ---%

A profound insight has been gained in a comparison between results
with full quantum interactions and classical truncations using the
Boltzmann equation for $\lambda\phi^4$ theory
\cite{Epelbaum:2015vxa}.  As explained in the last part in
\sref{sec:dense}, when $f\gg 1$, the cubic term $\sim f^3$ should be
dominant among the collision terms in the $2\leftrightarrow 2$
processes.  Such an approximation to truncate the collision terms to
keep $\sim f^3$ only and discard $\sim f^2$ should correspond to the
approximation employed in the classical statistical simulation.  It is
therefore intriguing to confirm numerically $\beta_{\rm eff}\to 0$
from the quantum Boltzmann equation and $\beta_{\rm eff}\to 2/3$ from
the classical truncations.  The practically important question is, in
particular, whether $\beta_{\rm eff}\simeq 0$ could be realized even
transiently or not in the classical approximation, and if the results
are affirmative, there may be still a good chance to utilize the
semi-classical approximation to resolve the isotropization problem.
The numerical results are summarized in an illustration in
\fref{fig:qboltzmann}.  It is a striking feature that the
semi-classical results monotonically converge to the so-called
classical attractor with $\beta_{\rm eff}=2/3$ and does not come close
to $\beta_{\rm eff}\sim 0$ at all.  As a matter of fact,
\fref{fig:qboltzmann} is an apparently different but equivalent
representation of the sketch in \fref{fig:kurkela}.

It is thus an urgent problem in theory to pursue for some
interpolating description used from the dilute to the dense regimes.
Especially in the dense regime, $f\gg 1$ does not hold for all the
momenta, and so the conventional methods are inadequate for the large
momentum regions.  So far, it remains as a tough open question how to
improve the theoretical formulation in the dense regime extrapolatably
toward the dilute regime.

%%%%%%%%%%   More on the Onset of Hydrodynamics   %%%%%%%%%%
\section{More on Onset of Hydrodynamics} 
\label{sec:hydrodynamics}

It would be a reasonable question to wonder what would happen if the
hydrodynamic equations are forcefully employed when $\PL\ll \PT$.  If
the usage of the hydrodynamic equations were legitimate even with
anisotropic pressures, we do not have to try to isotropize the system
using the plasma/glasma instabilities but simply switch to
hydrodynamics immediately after the glasma initial condition.  In an
ordinary sense, anisotropic pressures are accompanied by dissipative
terms and if the anisotropy is large such that the derivative
expansion can no longer be justified, we should not utilize the
hydrodynamic equations. Recently, however, a promising project for
resummed anisotropic hydrodynamic is ongoing under the name of the
aHydro \cite{Florkowski:2010cf,Martinez:2010sc}.  This section is
devoted to a brief summary of this interesting and still developing
subject.

%%%%%   Basics of Hydrodynamics   %%%%%
\subsection{Basics of Hydrodynamics}
Here, we would not attempt to elucidate the systematic derivation of
the hydrodynamic equations, but instead, we just take a quick look at
some basic expressions which are later necessary for the understanding
of the effect of anisotropic pressures.

We follow the discussions in \cite{Muronga:2003ta}.  The starting
point for hydrodynamics is the energy and the momentum conservation
laws expressed in terms of the energy-momentum tensor as
\begin{equation}
  \partial_\mu T^{\mu\nu} = 0\;.
\end{equation}
If there are some conserved charges such as the electric charge and
the baryon number, the continuity equation for such quantities should
be coupled, which we neglect in this subsection for simplicity.  The
fundamental variable for the hydrodynamic description is the velocity
vector $u^\mu$ which is normalized as $u_\mu u^\mu=1$.  Then, the
energy momentum tensor can be decomposed as
\begin{equation}
  T^{\mu\nu} = \varepsilon u^\mu u^\nu - (P+\Pi)\Delta^{\mu\nu}
  + W^\mu u^\nu + W^\nu u^\mu + \pi^{\mu\nu}\;,
\label{eq:Tmunu}
\end{equation}
where $\Delta^{\mu\nu}\equiv g^{\mu\nu}-u^\mu u^\nu$.  The physical
interpretation of each term is quite clear; $\varepsilon$ is the
energy density, $P+\Pi$ is the pressure, $W^\mu$ is the energy flow,
and $\pi^{\mu\nu}$ is the viscous stress tensor, which is defined by
$\pi^{\mu\nu}=T^{\langle\mu\nu\rangle}\equiv
\Delta^{\mu\nu}_{\;\;\alpha\beta} T^{\alpha\beta}$, where
\begin{equation}
  \Delta^{\mu\nu\alpha\beta} \equiv \frac{1}{2}(\Delta^{\mu\alpha}
  \Delta^{\nu\beta}+\Delta^{\nu\alpha}\Delta^{\mu\beta})
  -\frac{1}{3}\Delta^{\mu\nu}\Delta^{\alpha\beta}\;.
\end{equation}
It is easy to show that $\Delta^\mu_{\;\mu\alpha\beta}=0$, from which
$\pi^\mu_{\;\mu}=0$ follows.  Because $u_\mu\Delta^{\mu\nu}=0$ by
definition, we easily see $u_\mu\pi^{\mu\nu}=0$.  For the hydrodynamic
description we further need specify the choice of $u^\mu$;  in other
words, we are supposed to choose which conserved quantity flows with
$u^\mu$.  For the relativistic theory the energy current would be the
most convenient choice, so that $T^\mu_{\;\nu}u^\nu=\varepsilon u^\mu$
follows.  Such a frame in which $u^\mu$ represents the energy current
is often called the \textit{Landau frame}.  In this frame, $W^\mu=0$
simplifies the structure of the energy-momentum tensor a bit and we
should still need solve $\Pi$ and $\pi^{\mu\nu}$.

These continuity equations do not form a closed set of equations and
more unknown variables are contained in $\Pi$ and $\pi^{\mu\nu}$ than
equations even with an additional constraint from the equation of
state, $P=P(\varepsilon)$, given from thermodynamics.  To find an
explicit form of $\Pi$ and $\pi^{\mu\nu}$, a phenomenological argument
makes use of the 2nd-law of thermodynamics (see \cite{Hayata:2015lga}
for more field theoretical derivation of relativistic hydrodynamics),
i.e.\ $\partial_\mu s^\mu\ge 0$ for a given entropy current $s^\mu$.
The entropy current could be expressed as
\begin{equation}
  s^\mu = \frac{1}{T}T^{\mu\nu}u_\nu + \frac{p}{T}u^\mu + Q^\mu\;,
\end{equation}
where the last term $Q^\mu$ represents 2nd-order dissipative terms
that are not taken into account in the 1st-order viscous
hydrodynamics.  In the 1st-order theory the divergence of the entropy
current turns out to be
\begin{equation}
  T\partial_\mu s^\mu = -\Pi\cdot \nabla_\mu u^\mu
  + \pi_{\mu\nu}\cdot \partial^{\langle \mu} u^{\nu\rangle}\;,
\end{equation}
where $\nabla_\mu \equiv \Delta_{\mu\nu}\partial^\nu$.  To satisfy
$\partial_\mu s^\mu\ge 0$, it would be sufficient to require that
$\partial_\mu s^\mu$ is a sum of squared quantities, which immediately
leads to
\begin{equation}
  \Pi = -\zeta \nabla_\mu u^\mu\;,\qquad
  \pi^{\mu\nu} = 2\eta\,\partial^{\langle\mu} u^{\nu\rangle}
\label{eq:1st}
\end{equation}
with some positive coefficients, $\zeta$ and $\eta$, which are called
the bulk and the shear viscosities, respectively.  These transport
coefficients should be given as inputs (from the linear-response
theory, for example) to viscous hydrodynamics.

This framework of the 1st-order hydrodynamic equations has a serious
flaw violating the causality.  The problem can be cure by the
2nd-order theory that incorporates neglected $Q^\mu$;  we can obtain
$Q^\mu$ in terms of $u^\mu$, $\Pi$, and $\pi^{\mu\nu}$, and the
condition for $\partial_\mu s^\mu$ involves their time derivatives, so
that \eref{eq:1st} should be replaced with the equations of motion for
$\Pi$ and $\pi^{\mu\nu}$ with more transport coefficients, namely, the
relaxation times, $\tau_\Pi$ for $\Pi$ and $\tau_\pi$ for
$\pi^{\mu\nu}$.  If these relaxation times are large enough, there is
no problem to violate the causality.

It would be useful to write some of the explicit equations down here
in a special case corresponding to the heavy-ion collision as
addressed in \cite{Muronga:2001zk}.  Let us focus on the
boost-invariant and conformal case and then we can drop the bulk
viscosity effect.  We can then simplify the 2nd-order viscous
equations significantly as
\begin{equation}
  \frac{\rmd \varepsilon}{\rmd\tau}
  + \frac{\varepsilon + P}{\tau} = \frac{\Phi}{\tau}\;,
\label{eq:e_PL2}
\end{equation}
where $\Phi\equiv \pi^{00}-\pi^{zz}$.  (In \cite{Muronga:2001zk} there
was a factor $2/3$ error, which was corrected in an erratum.)  Also,
there is an equation to determine $\Phi$ that reads:
\begin{equation}
  \frac{\rmd\Phi}{\rmd\tau} = -\frac{\Phi}{\tau_\pi}
  - \frac{\Phi}{2}\biggl[\frac{1}{\tau} + \frac{T}{\beta_2}
    \frac{\rmd}{\rmd \tau}\biggl(\frac{\beta_2}{T}\biggr)\biggr]
  + \frac{2}{3\beta_2 \tau}
\end{equation}
with $\beta_2=\tau_\pi/(2\eta)$.  This equation can be further
simplified with the conformality assumption in which
$\beta_2\propto T^{-4}$ and we can use $T\propto \tau^{-1/3}$ for the
term proportional to $\Phi/\beta_2$ in the construction of the
2nd-order theory.  Then, the $\tau$ derivative gives a factor $5/3$,
and we eventually have,
\begin{equation}
  \frac{\rmd\Phi}{\rmd\tau} = -\frac{\Phi}{\tau_\pi}
  - \frac{4\Phi}{3\tau} + \frac{4\eta}{3\tau_\pi \tau}\;.
\label{eq:Phi}
\end{equation}
These are simple but useful equations for the benchmark purpose.  Any
extension of the hydrodynamic equations with resummation with respect
to anisotropy should reproduce them once expanded in terms of small
anisotropy, which will be checked later.

%%%%%   Dissipative Terms and Anisotropy   %%%%%
\subsection{Dissipative Terms and Anisotropy}
The ordinary hydrodynamic equations have only single $P$ assuming
isotropy.  The tensor decomposition in \eref{eq:Tmunu} suggests that
$\PL\neq\PT$ could be related to some dissipative terms from
$\pi^{\mu\nu}$, and this is indeed true.  Therefore, if we improve
hydrodynamics from the 1st-order theory to the 2nd-order theory, we
may have a better situation to treat a larger deviation of
$\PL/\PT\neq 1$, but eventually, we need reorganize the derivative
expansion shifting the expansion reference point.  In general
circumstances such reorganization is a tough problem but the kinetic
equation can provide us with a useful guide to the right path.

%%%   Hydrodynamic interpretation of anisotropy   %%%
\subsubsection{Hydrodynamic interpretation of anisotropy}
Interestingly, \eref{eq:e_PL2} is a hydrodynamic counterpart of
\eref{eq:e_PL}.  We note that \eref{eq:e_PL} is an exact relation
regardless of hydrodynamics, and this means that, by taking a
difference between \eref{eq:e_PL} and \eref{eq:e_PL2}, we have
$P-\Phi=\PL$.  We here assume that $P$ appearing in the previous
subsection is identifiable to an average; $(2\PT+\PL)/3$.  Also, in
the 1st-order theory, $\tau_\pi\to0$, and this implies that two terms
in \eref{eq:Phi} should cancel, leading to a relation;
$\Phi=4\eta/(3\tau)$ \cite{Muronga:2001zk}.  Combining these
relations, we finally reach the following;
\begin{equation}
  \PT - \PL = \frac{3}{2}\Phi = \frac{2\eta}{\tau}\;.
\label{eq:peta}
\end{equation}
This clearly shows that $\PL/\PT\neq 1$ should be accompanied by a
shear viscosity, and if it substantially remains constant at later
time, it would favor a large value of the shear viscosity, while
$\eta$ could be small for small $\tau$.  In fact, in the classical
statistical simulation with $\PL/\PT\sim 0.6$ the viscosity to the
entropy density ratio was estimated in this way and the result was
$\eta/s\sim 0.3$ \cite{Gelis:2013rba}, which is much smaller than the
perturbative estimate (for the coupling $g=0.5$ that was adopted in
the numerical simulation).

%---   figure   ---%
\begin{figure}
 \begin{center}
 \includegraphics[width=0.65\textwidth]{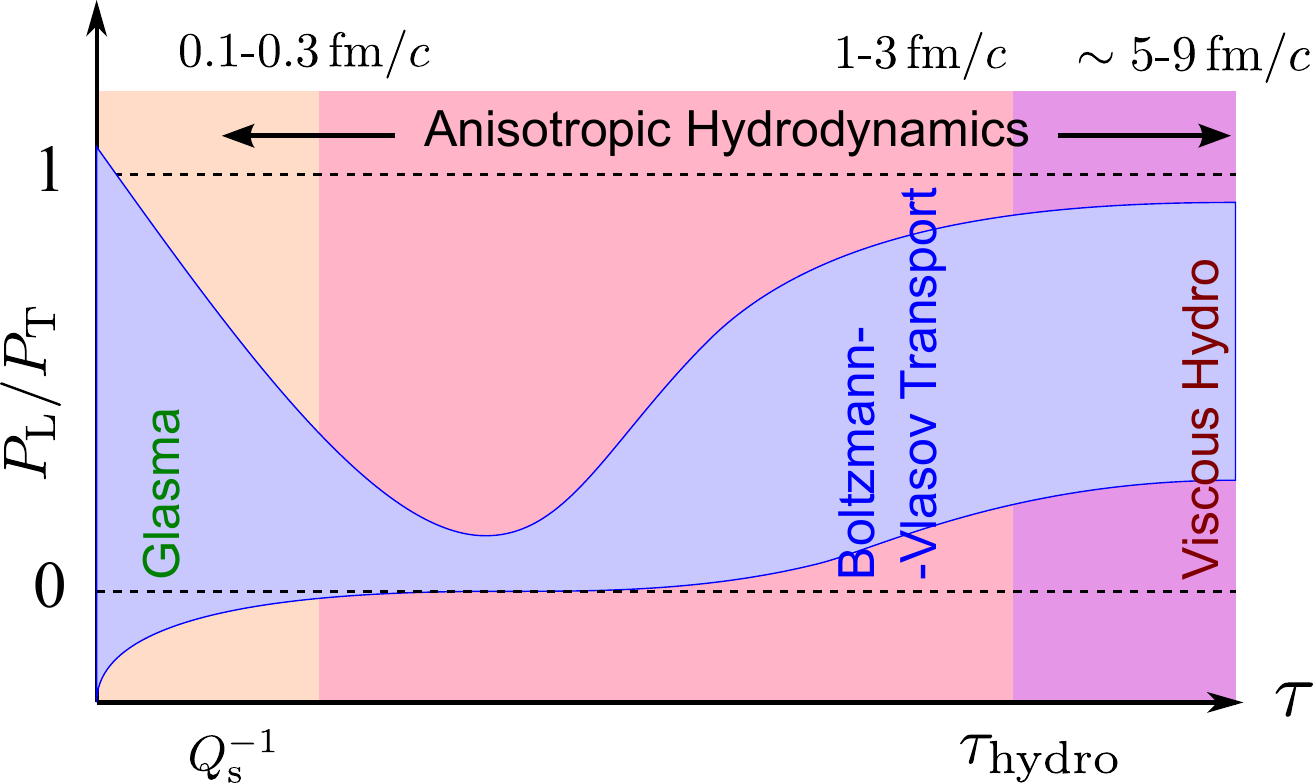}
 \end{center}
 \caption{Schematic picture of the evolution of $\PL/\PT$ and the
   hydrodynamization time $\tau_{\rm hydro}$ which can be taken to be
   smaller once the hydrodynamic equations are augmented with large
   anisotropic effects.  Figure adapted from a talk by Strickland.}
  \label{fig:strickland}
\end{figure}
%---   figure   ---%

We can equivalently rewrite \eref{eq:peta}, using
$s\sim \partial P/\partial T\sim 4P/T$, into a form of the ratio as
follows:
\begin{equation}
  \frac{\PL}{\PT} = \frac{3\tau T-16(\eta/s)}{3\tau T+8(\eta/s)}\;,
\label{eq:peta2}
\end{equation}
which implies $\PL/\PT\approx 0.5$ for initial $\tau_0\sim 0.5\fm/c$
and $T_0\sim 0.4\GeV$, if $\eta/s\sim 1/(4\pi)$ is assumed, at RHIC
energy.  The reduction of $\PL/\PT$ suggested by \eref{eq:peta} or
\eref{eq:peta2} gives us a motive to pursue for an improved
hydrodynamic formulation that can incorporate $\PL/\PT\ll 1$.  The
schematic (desirable) picture is illustrated in \fref{fig:strickland}
which is adapted from Strickland's picture.  It is usually believed
that at the hydrodynamization time $\tau_{\rm hydro}$ we should switch
the theoretical description from the kinetic equations
(Boltzmann-Vlasov equations) to the hydrodynamic ones.  If we use
higher-order viscous hydrodynamics, we could take $\tau_{\rm hydro}$
to be smaller, and ideally, if we have an optimized resummed scheme,
it may be not really hopeless to anticipate that such resummed
hydrodynamics can entirely cover the time evolution superseding the
kinetic equations at all.

%%%   Resummed anisotropic hydrodynamics   %%%
\subsubsection{Resummed anisotropic hydrodynamics}
A pioneering work for anisotropic hydrodynamics is found in
\cite{Florkowski:2008ag} by the Krakow group, in which the continuity
equation for the entropy current was assumed.  Soon later, the
formulation was refined in \cite{Florkowski:2009sw,Florkowski:2010cf}
where the distribution function was considered with the entropy
conservation replaced by the particle number conservation.  Around the
same timing, independently in \cite{Martinez:2009mf} by the Frankfurt
group, the anisotropic parameter, $\Delta\equiv\PT/\PL-1$, was studied
with 2nd-order conformal hydrodynamics.  There, also the relation
between $\Delta$ and the anisotropic parameter appearing in
\eref{eq:anisotropic} was clarified in the Appendix.  A systematic
presentation in terms of the distribution function was given in
\cite{Martinez:2010sc}.  Then, as argued in \cite{Ryblewski:2010ch},
it turned out that ADHYDRO (highly-anisotropic and
strongly-dissipative hydrodynamics) of \cite{Florkowski:2010cf} and
aHydro of \cite{Martinez:2010sc} have equivalent physical contents
microscopically.

The hydrodynamic equations are the equations of motion for the
energy-momentum tensor, which can be inferred from the kinetic
equations.  Then, the clearest strategy is as follows; using the
kinetic equations with an anisotropy parameter as introduced in
\eref{eq:anisotropic}, we can systematically derive the anisotropic
hydrodynamic equations.  Unlike \eref{eq:anisotropic} the anisotropy
parameter should be a function of $\tau$, which is denoted here by
$\xi(\tau)$, and the distribution function is then parametrized as
\begin{equation}
  f(\bk,\tau) = f_{\rm iso}\Bigl(\Lambda^{-1}(\tau)
  \sqrt{\bk^2+\xi(\tau)k_z^2}\Bigr) \;.
\label{eq:fiso}
\end{equation}
In the $\xi\to0$ limit the distribution is reduced to an isotropic
one (here, note that $\bk^2=k_\perp^2+k_z^2$), and the distribution is
prolate and oblate deformed, respectively, for $-1<\xi<0$ and
$\xi>0$.

As explained in \sref{sec:dilute} the simplest approximation for the
Boltzmann collision term is the RTA, with which the Boltzmann equation
reads:
\begin{equation}
  k^\alpha\partial_\alpha f = -k_\mu u^\mu \Gamma\bigl[
    f(t,z,\bk)-f_{\rm eq}(t,z,\bk,T(\tau))\bigr]\;.
\label{eq:aBoltzmann}
\end{equation}
We could, in principle, deal with the above equations to solve
$\xi(\tau)$ and $\Lambda(\tau)$.  Instead, we can rewrite
\eref{eq:aBoltzmann} into a form similar to the hydrodynamic
equations by taking the moments of \eref{eq:aBoltzmann}.  The 0th
moment of \eref{eq:aBoltzmann} leads to
\begin{equation}
  \frac{\partial_\tau\xi}{1+\xi} - \frac{2}{\tau}
  - 6\partial_\tau \log\Lambda
  = 2\Gamma\Bigl[ 1-\calR^{3/4}(\xi)\sqrt{1+\xi} \Bigr]\;,
\end{equation}
where $\calR(\xi)\equiv \frac{1}{2}
[1/(1+\xi)+(\arctan\sqrt{\xi})/\sqrt{\xi}]$.  Because there are two
variables, $\xi(\tau)$ and $\Lambda(\tau)$, we need one more
equation from the 1st moment of \eref{eq:aBoltzmann}, i.e.
\begin{equation}
  \frac{\calR'(\xi)}{\calR(\xi)}\partial_\tau\xi
  + 4\partial_\tau \log\Lambda
  = \frac{1}{\tau}\biggl[ \frac{1}{\xi(1+\xi)\calR(\xi)}
    - \frac{1}{\xi}-1 \biggr]\;.
\end{equation}
When the anisotropy parameter is small; $\xi\ll 1$, these equations
are equivalent to \eref{eq:e_PL2} and \eref{eq:Phi} once the
linearized solution $\xi\approx (45/8)\Phi/\varepsilon$ and
$\Gamma=2/\tau_\pi$ are used.  Therefore, these coupled equations for
$\xi(\tau)$ and $\Lambda(\tau)$ are to be regarded as an anisotropic
upgrade of the 2nd-order viscous hydrodynamics, i.e., an aHydro
formulation.  Remarkably, these equations in the leading-order aHydro
are capable of capturing the expected features in
\fref{fig:strickland} including the region where $\PL/\PT\lesssim 0.1$
or even smaller.

Later, this formalism has been extended including the transverse
dynamics, the next-to-leading order fluctuations \cite{Bazow:2013ifa},
and also the mass effects that breaks conformal symmetry and thus
induces a finite bulk viscosity \cite{Nopoush:2014pfa}, which was also
addressed in \cite{Tinti:2014yya}.  The interesting phenomenological
implication is that there could be a difference in the temperature
slopes as well as in the pressures.  In \eref{eq:fiso} $\Lambda(\tau)$
should correspond to the transverse temperature $T_{\rm T}$, and in
the anisotropic limit where $\xi\gg 1$, the longitudinal temperature
should be $T_{\rm L}=\Lambda(\tau)/\xi(\tau)\ll T_{\rm T}$.

The problem in this approach is an ambiguity in specifying the
equation of state for anisotropic matter.  As long as the underlying
microscopic dynamics is known for a given distribution function with
an anisotropy parameter $\xi$, no such problem arises manifestly.  If
it is ultimately intended to get rid of the kinetic description at
all, the relation between the equation of state and $\xi$ should be a
part of unclear assumption.

%%%%%%%%%%   More on Spectral Cascade   %%%%%%%%%%
\section{More on Spectral Cascade}
\label{sec:thermal}

The isotropization quantified by $\PL$ and $\PT$ in
\sref{sec:isotropization} and the hydrodynamization in
\sref{sec:hydrodynamics} are both integrated properties of matter, and
in this section, we will discuss more differential properties, namely,
the real-time evolution of the particle distribution as a function of
the momentum.  We already flashed some scaling arguments in
\sref{sec:isotropization} and in this section we will specifically
look at a speculative scenario suggested from the CGC initial
condition (and see also \cite{Kurkela:2011ti} for similar analysis).
The question is the following;  the CGC state is saturated with gluons
and such abundant gluons seem to be not really accommodated in a
thermal distribution function.  This observation naturally leads to an
idea of a transient formation of the gluonic BEC during the
thermalization processes, or a generic picture of dynamical BEC
formation with an overpopulated initial condition.

%%%%%   CGC-based Scenario   %%%%%
\subsection{CGC-based Scenario and the Bose-Einstein Condensate}
\label{sec:BEC}

\begin{figure}
 \begin{center}
 \includegraphics[width=0.7\textwidth]{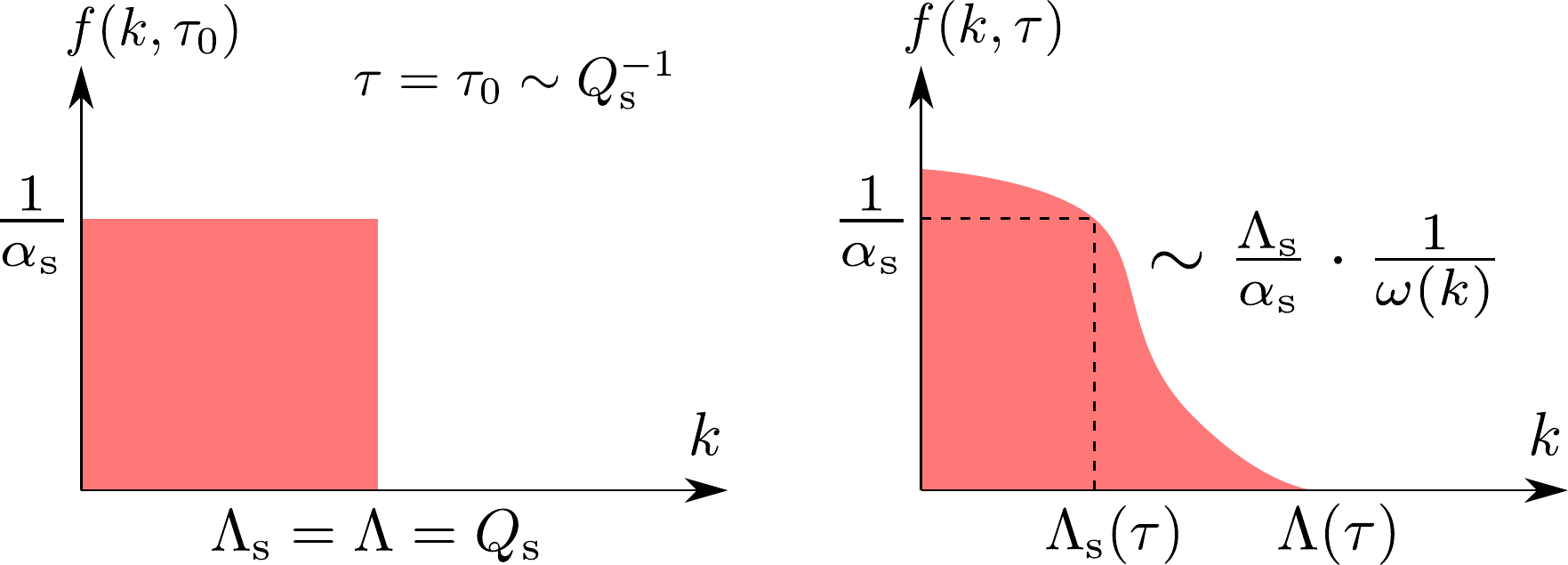}
 \end{center}
 \caption{Schematic picture of the real-time evolution of the
   distribution function from the CGC initial condition (left) to late
   time profile (right).}
  \label{fig:BEC}
\end{figure}

It was recognized since \cite{Blaizot:2011xf} that the overpopulated
initial condition could generally have peculiar dynamics, which may be
the case for the heavy-ion collisions.  In this scenario there are two
characteristic scales; an IR scale $\Lambda_{\rm s}(\tau)$ and a UV
scale $\Lambda(\tau)$.  At the initial time $\tau=\tau_0$, it is
assumed that $\Lambda_{\rm s}(\tau_0)=\Lambda(\tau_0)\sim \Qs$ and the
shape of the distribution function is approximated as
$f(\bk)\sim \alphas^{-1}$ for $|\bk|<\Qs$ and $f(\bk)\sim 0$ for
$|\bk|>\Qs$ as sketched in the left of \fref{fig:BEC}.  This is a
simplified version of the CGC initial condition that correctly
captures the qualitatively essential features.  The important
observation is that, as time goes, $\Lambda_{\rm s}(\tau)$ decreases
and $\Lambda(\tau)$ increases and the thermal distribution arises in
the window between $\Lambda_{\rm s}(\tau)$ and $\Lambda(\tau)$ as
depicted in the right of \fref{fig:BEC}, which is parametrized as
\begin{equation}
  f(\bk) \sim \frac{\Lambda_{\rm s}}{\alphas}\cdot
  \frac{1}{\omega(\bk)} \qquad
  \mbox{for~~ $\Lambda_{\rm s} < |\bk| < \Lambda$}\;.
\label{eq:BECtherm}
\end{equation}

%---   figure   ---%
%---   figure   ---%

An interesting question is to determine the parametric dependence of
$\Lambda_{\rm s}(\tau)$ and $\Lambda(\tau)$.  According to the
arguments in \cite{Blaizot:2011xf} they should parametrically depend
on $\tau/\tau_0$ as
\begin{equation}
  \Lambda_{\rm s}(\tau) \sim \Qs\biggl(\frac{\tau}{\tau_0}\biggr)^{-3/7}\;,
  \qquad
  \Lambda(\tau) \sim \Qs\biggl(\frac{\tau}{\tau_0}\biggr)^{1/7}\;.
\label{eq:BECscale}
\end{equation}
The derivation of the above results is as follows.  The total energy
$\sim \Lambda_{\rm s}\Lambda^3$ should be conserved, for which the
energy is dominated by a contribution from the window in
\eref{eq:BECtherm} with the UV cutoff by $\Lambda$.  (The contribution
from $|\bk|<\Lambda_{\rm s}$ is suppressed by
$(\Lambda_{\rm s}/\Lambda)^2$.)  Obviously \eref{eq:BECscale}
satisfies this condition.  Another condition comes from the typical
time scale $\tau_{\rm scat}$ of the collision, which can be estimated
from the scattering amplitude squared.  For example of the
$2\leftrightarrow 2$ scattering,
$\tau_{\rm scat}^{-1}\sim \alphas^2 (\Lambda_{\rm s}/\alphas)^2
\Lambda^{-1} \sim \Lambda_{\rm s}^2 \Lambda^{-1}$ with
$\Lambda_{\rm s}/\alphas$ appearing from the distribution.  With the
concrete form of the kinetic equation, it is concluded
\cite{Blaizot:2011xf} that
$\tau\sim \tau_{\rm scat}\sim \Lambda_{\rm s}^{-2}\Lambda$, from which
\eref{eq:BECscale} follows.

If the thermalization occurs, there should be a clear scale separation
between the hard and the soft components by the difference in terms of
the strong coupling constant, i.e., the thermalization condition could
be chosen as $\Lambda\sim T$ and $\Lambda_{\rm s}\sim \alphas T$,
which implies together with \eref{eq:BECscale} that the thermalization
time is estimated as
\begin{equation}
  \Qs\tau_{\rm th} \sim \alphas^{-7/4}\;,\qquad
  T/\Qs \sim \alphas^{-1/4}\;,
\end{equation}
which makes a sharp contrast to \eref{eq:thermtime_bottom} in the
bottom-up thermalization scenario.

The above-mentioned arguments lead us to an interesting speculation.
The initially adopted distribution of gluons (left of \fref{fig:BEC})
gives the number of gluons as
$n_0 = n_{\rm g}(\tau_0) \sim \Qs^3/\alphas$, which evolves as
$n_{\rm g}(\tau) \sim \Lambda^2 \Lambda_{\rm s}/\alphas$ in later time
(right of \fref{fig:BEC}) and at $\tau=\tau_{\rm th}$ the thermal
gluon number is naturally
$n_{\rm g}(\tau_{\rm th}) \sim T^3 \sim \alphas^{-3/4}\Qs^3$.  If the
elastic scatterings are dominant over the inelastic ones (which is the
case for $f\gg 1$), the gluon number can be regarded as a conserved
quantity.  Then, the discrepancy between $n_0$ and
$n_{\rm g}(\tau_{\rm th})$ should be sent to a condensate at zero
mode, that is, the gluonic BEC should develop so that the condensed
gluon number,
\begin{equation}
  n_{\rm c} = n_0-n_{\rm g}(\tau_{\rm th}) \sim
  \bigl(1-\alphas^{1/4}\bigr) \frac{\Qs^3}{\alphas} \;,
\label{eq:BECestimate}
\end{equation}
can compensate for the mismatch.  For a realistic situation with
$\alphas\sim 0.3$ for instance, \eref{eq:BECestimate} means that about
$26\%$ of initially saturated gluons should fall into a BEC when
thermalization is achieved.  Of course, this estimate is based on
quite optimistic simplification.

%%%%%   Dynamical Evolution of the Spectral Cascade   %%%%%
\subsection{Dynamical Evolution of the Spectral Cascade}
Let us now consider the evolution of the whole spectral shape.  It is
a common phenomenon that a power-law spectrum appears as a steady
solution out of equilibrium.  We can find the power index from the
kinetic equation arguments.

%%%   Energy cascade vs. particle cascade   %%%
\subsubsection{Energy cascade vs.\ particle cascade}
To consider the dynamical evolution of the power-law spectrum in QCD,
it would be very helpful to gain more general understanding for the
spectrum associated with the wave turbulence.  Wave turbulence is a
phenomenon that occurs in random non-linear waves such as
gravity-capillary waves and should be clearly distinguished from the
hydrodynamic turbulence.

The theoretical setup is as follows;  the system has constant energy
pumping at small $k_-$ and energy damping at larger $k_+$ and we can
use the kinetic equation in an interval, $k_-\ll k\ll k_+$ to describe
the wave turbulence to find an index $\nu$ of the Kolmogorov-Zakharov
(KZ) spectrum that characterizes the distribution function as
\begin{equation}
  f(\bk) \sim \frac{1}{k^\nu}\;.
\end{equation}
For a given collision kernel of, for example, $2\leftrightarrow 2$
scattering, we can  identify the value of $\nu$, and there may
sometimes be multiple solutions; one corresponds to the
\textit{energy cascade} and the other corresponds to the
\textit{particle cascade}.

It is easier to find $\nu$ using the continuity equation rather than
solving the Boltzmann equation.  Let us consider a scattering process
involving $p$ particles (waves), and suppose the following scaling
properties for the energy dispersion relation and the scattering
amplitude as
\begin{equation}
  \omega(\mu\bk) = \mu^\alpha\omega(\bk)\;,\qquad
  \calM(\mu\bk_1,\mu\bk_2,\dots) = \mu^\beta
  \calM(\bk_1,\bk_2,\dots)\;.
\end{equation}
We first consider the energy cascade, as a result of which the energy
spectrum becomes time independent.  In terms of the distribution
function the one-dimensional energy spectrum is written down as
\begin{equation}
  E_k(t) = \int \rmd\Omega\, k^{d-1} \omega(\bk)f(\bk,t)\;.
\end{equation}
Here, $d$ represents the number of spatial dimensions.  The energy
conservation is expressed by the energy continuity equation:
\begin{equation}
  \frac{\partial E_k}{\partial t}
  + \frac{\partial \varepsilon_k}{\partial k} = 0\;,
\label{eq:Ek}
\end{equation}
where the energy flow flux is given by
\begin{equation}
  \varepsilon_k = \int_{k_+}^k \rmd q\int \rmd\Omega\,q^{d-1}
  \omega(\bq)\,
  \biggl(\frac{\partial f}{\partial t}\biggr)_{\rm coll}
  \sim k^{(p-1)d -p\alpha +2\beta-\nu(p-1)} \;.
\label{eq:ek}
\end{equation}
When the system reaches a steady state in terms of the energy,
$\partial E_k/\partial t=0$ is realized, and then $\varepsilon_k$ must
be independent of $k$.  From this condition of
$\varepsilon_k\sim k^0$ (not the zeroth component but the zeroth power
of $k$), we can get to the KZ spectrum index from the energy cascade:
\begin{equation}
  \nu_{\rm E} = d + \frac{2\beta-p\alpha}{p-1}\;.
\label{eq:nu_e}
\end{equation}
For $2\leftrightarrow 2$ scattering in a (3+1)-dimensional massless
scalar theory, $d=3$, $\alpha=1$, and $\beta=0$ and thus we have
$\nu_{\rm E}=5/3$.  We note that the above mentioned derivation of the
index for the KZ spectrum is quite analogous to the famous Kolmogorov
spectrum $E_k\sim k^{-5/3}$ for hydrodynamic turbulence for which an
energy flow flux is assumed to be $k$ independent.  Differently from
the Kolmogorov spectrum that is fixed by the dimensional analysis,
$\nu_{\rm E}$ is sensitive to the structure of microscopic
interactions as seen in \eref{eq:nu_e}.  For example, the capillary
waves on deep water has $\nu_{\rm E}=17/4$ (Zakharov-Filonenko
spectrum), the acoustic turbulence $\nu_{\rm E}=9/2$ (Zakharov-Sagdeev
spectrum), etc.  Another branch of the solution belongs to the
particle cascade.  In this case we should consider the particle number
continuity equation:
\begin{equation}
  \frac{\partial N_k}{\partial t} + \frac{\partial \mu_k}{\partial k}
  = 0\;,
\end{equation}
where $N_k$ is the one-dimensional particle number and $\mu_k$ is the
particle flow flux, which are defined in the same was as the energy
cascade with $\omega(\bk)$ removed from \eref{eq:Ek} and
\eref{eq:ek}, that after all leads to
\begin{equation}
  \nu_{\rm N} = d + \frac{2\beta - (p+1)\alpha}{p-1}\;.
\label{eq:nu_n}
\end{equation}
Hence, the particle cascade results in the KZ spectrum with
$f\sim k^{-4/3}$ for $2\leftrightarrow 2$ scattering in a massless
theory, while it was $f\sim k^{-5/3}$ from the energy cascade.

Generally speaking, the particle cascade occurs in a direction toward
smaller $k$ and the energy cascade toward larger $k$.  Let us recall
that we consider the spectral cascade from $k_-$ to $k_+$ (where
$k_-\ll k_+$) and introduce the one-particle energy $\omega_\pm$
(where $\omega_-\ll \omega_+$) and the particle flow $\mu_\pm$ at
$k=k_\pm$.  Then, the total quantities are $\mu=\mu_++\mu_-$ and
$\varepsilon=\omega_+\mu_++\omega_-\mu_-$, which
leads to
\begin{equation}
  \mu \simeq \mu_-\;,\qquad
  \varepsilon \simeq \omega_+\mu_+\;.
\label{eq:muepsilon}
\end{equation}
This analysis with \eref{eq:muepsilon} implies that the dynamical
evolution of the spectrum generally consists of the direct energy
cascade (from smaller $k$ to $k_+$) and the inverse particle cascade
(from larger $k$ to $k_-$).

%%%   Scenario with non-thermal fixed-point   %%%
\subsubsection{Scenario with non-thermal fixed-point}

As explained in \sref{sec:BEC} it is likely that a certain amount of
particles fall into the zero mode if the initial state is
overpopulated and the particle number is approximately conserved.
Once this situation happens the expected spectrum should be changed
because a condensate allows for $1\leftrightarrow 2$ scattering.  This
means that we should plug $p=3$ (keeping $\beta=0$ because a scalar
condensate does not scale with the momentum) into \eref{eq:nu_e} and
\eref{eq:nu_n} to find,
\begin{equation}
  \nu_{\rm E} = \frac{3}{2}\;,\qquad
  \nu_{\rm N} = 1\;.
\end{equation}
In this case, there is another possibility, which is referred to as
the non-thermal fixed point \cite{Berges:2008wm}.  We note that the
indices identified with the Boltzmann equations should be valid in
the perturbative regime only and the collision terms must involve
infinite Feynman diagrams once the interaction goes beyond the
perturbatively manageable range.  Such non-perturbative treatments are
mandatory once the coupling constant ($\lambda$ in a scalar theory)
and/or the distribution function $f$ become large.  Actually, if
$f\gtrsim 1/\lambda$ in a scalar theory or $f\gtrsim 1/\alphas$ in the
saturated regime of QCD, there should be more contributions from terms
with more $f$'s and a simple counting in \eref{eq:ek} breaks down.

Then, we should switch the theoretical tool from the perturbative
Boltzmann equation to the non-perturbative Dyson-Schwinger (or
Kadanoff-Baym) equation to look for the scaling solution.  This
question was addressed in \cite{Berges:2008wm} within the framework of
the 2PI large-$N$ expansion of $O(N)$ scalar theory with a
condensation field.  The analytical solution is either $\nu=4$ or $5$
and the numerical simulation favors $\nu=4$, and it has been concluded
in \cite{Berges:2008wm} that
\begin{equation}
  \nu_{\tiny \mbox{non-thermal}} = 4\;.
\end{equation}
A similar analysis by means of the Dyson-Schwinger equation has been
performed in a pure Yang-Mills theory in \cite{Carrington:2010sz}, and
it was found that $\nu=2$ and $5$ and even intermediate values might
be possible.

%---   figure   ---%
\begin{figure}
 \begin{center}
 \includegraphics[width=0.6\textwidth]{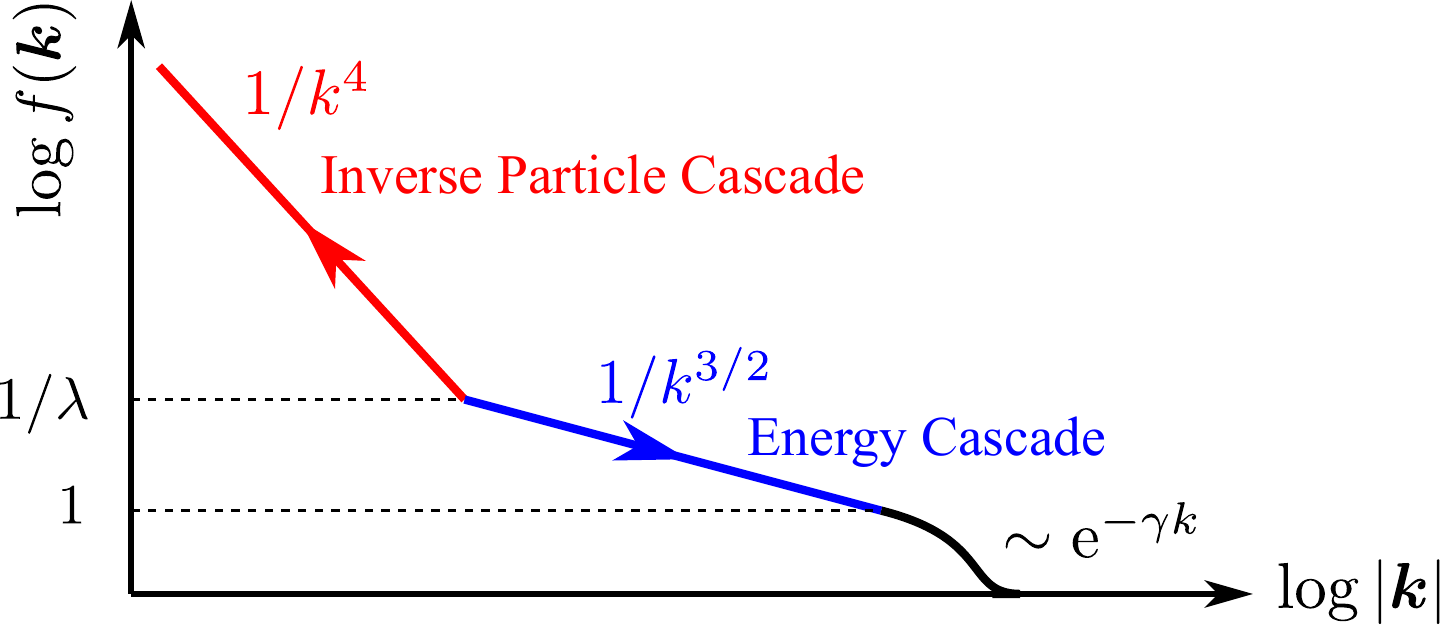}
 \end{center}
 \caption{Speculated distribution spectrum with a condensate at zero
   mode (without expansion).  Figure is adapted from
   \cite{Berges:2012us}.}
  \label{fig:nonthermal}
\end{figure}
%---   figure   ---%

Later, the scenario has been summarized as a diagram in
\fref{fig:nonthermal} according to Berges, and this has been confirmed
by the classical statistical simulation in \cite{Berges:2012us}.  When
$f(\bk)\lesssim 1$ quantum fluctuations dissipate high momentum modes
into heat and the tail behaves as $f \sim \rme^{-\gamma k}$.  In the
regime where $1\ll f(\bk)\ll 1/\lambda$, the kinetic theory works and,
as we already saw in the previous subsection, the direct energy
cascade is dominant for the flow to larger $k$, leading to
$f \sim 1/k^{3/2}$.  For small momenta the inverse particle cascade is
attributed to the BEC formation, but in this regime with
$f(\bk)\gg 1/\lambda$ the perturbative kinetic equation is no longer
useful.  Thus, instead of $\nu=1$, the index of the non-thermal fixed
point, $\nu=4$, should be the expected answer.

It is of course the most interesting question what would be the answer
for the situation relevant for the heavy-ion collision physics.  It is
still unclear whether a gluonic BEC could be formed in the classical
statistical simulation \cite{Berges:2012ev} (see also
\cite{Gasenzer:2013era} for analogous Higgs models).  It is even more
non-trivial how the scenario should be modified by the expansion
effect.  So far, all the discussions were focused on the massless case
only, and the mass effect was recently analyzed in
\cite{Blaizot:2015wga} in the case without expansion, and no
qualitative difference was found between the massless and the massive
cases.

%%%%%%%%%%   Further Topics   %%%%%%%%%%
\section{Further Topics}
There are many important topics that could not be covered by this
review due to limitation of the author's ability.  As a final remark,
here, we would not try to name all of such uncovered topics, but we
shall pay our attention to a particular problem, that is, the quark
production in the very early stage in the heavy-ion collision.  The
pictorial view in \fref{fig:glasma} leads us to the theoretical
formulation of the particle production according to the well-known
Schwinger mechanism.  In general, non-perturbatively, the production
amplitude is given by a Bogoliubov coefficient associated with the
basis transformation that connects quark states in the asymptotic past
and in the asymptotic future.  The Bogoliubov coefficient encompasses
all the perturbative processes, which can be easily confirmed by
perturbative expansion of the coefficient in terms of the coupling
constant.

The numerical simulation on top of the glasma configurations already
exists \cite{Gelis:2005pb} and it claims that most of quarks are
produced in a time scale $\lesssim \Qs^{-1}$.  In fact, in the limit
of infinitely thin nuclear sheets at high energies, the quark
production dominantly occurs at the light-cone singularity where the
color sources propagate, and thus, even at $\tau=0^+$, more than a
half of total quarks are produced instantly.  In the context of the
isotropization that we put our emphasis on in this review, the
produced quarks would hardly change the qualitative properties such as
the isotropization and the universal scaling exponents even though the
backreaction from the quark to the gauge sectors is taken into account
\cite{Gelfand:2016prm}.  Nevertheless, the problem of quark production
would spice our problem with a new physics opportunity, i.e., the
\textit{chirality}.

The reason why the chirality is such a special character of matter is
that it couples with the quantum anomaly and the QCD $\theta$-vacuum
structure.  Fortunately, the heavy-ion collision is an ideal
environment for such a study to explore anomaly-induced novel
phenomena;  we have the production of (almost) chiral quarks, and we
also have an experimental probe, that is, a strong magnetic field.
The coupling between the chirality and the magnetic field would
generate topological currents from the chiral magnetic effect, the
chiral separation effect, the chiral vortical effect, etc.  A full
explanation of those chiral topological effects requires another 30
pages review, and we would not go into further technical details on
this.  Interested readers can consult a recent status summary
\cite{Liao:2016diz}.

%---   figure   ---%
\begin{figure}
 \begin{center}
 \includegraphics[width=0.3\textwidth]{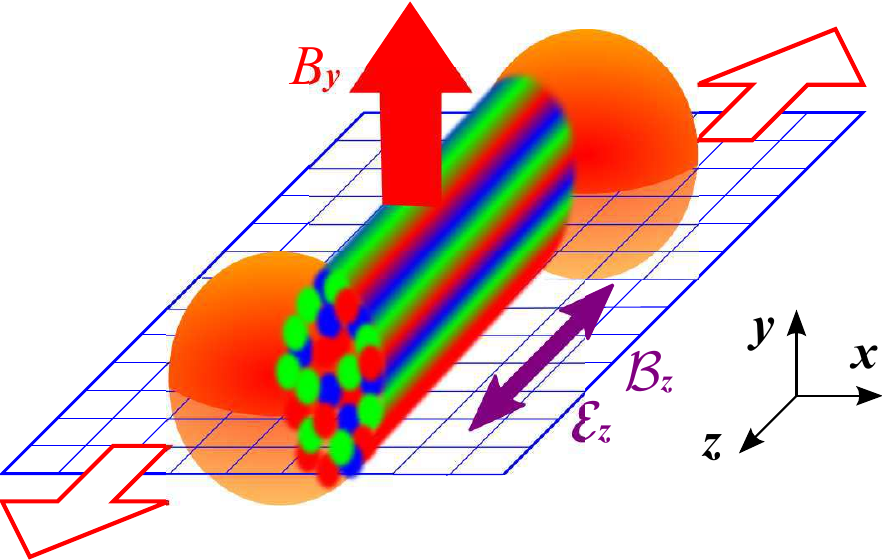}
 \end{center}
 \caption{Particle production on top of color flux tubes in the
   presence of magnetic field.  Figure taken from
   \cite{Fukushima:2010vw}.}
  \label{fig:cme}
\end{figure}
%---   figure   ---%

Most importantly, the color flux tube structure in the glasma initial
condition accommodates a bunch of domains with parallel color electric
and magnetic fields.  Because the electric field is a vector and the
magnetic field is an axial vector, the inner product of them is
parity-odd (and charge-parity-odd too).  Thus, the initial condition
in the heavy-ion collision has large topological fluctuations (see
\cite{Mace:2016svc} for a recent simulation to quantify this effect).
Naturally, as sketched in \fref{fig:cme}, the momentum distribution
$f(\bk)$ of produced quarks must exhibit some anisotropy under the
influence of external magnetic field.  Indeed, such skewed $f(\bk)$
was observed in an idealized simulation with homogeneous Abelian
fields \cite{Fukushima:2015tza}.  Efforts along these directions
should be appreciated, for the lifetime of the magnetic field is known
to be as short as comparable to $\sim \Qs^{-1}$.  Anomalous
hydrodynamics and chiral kinetic theory should be enumerated as
outstanding theoretical developments inspired by the chiral
topological effects.  However, they both need an initial condition for
any practical application, and the initial condition should be given
at the glasma time scale.

As stated in the very beginning of this review, early thermalization
is the last and greatest unsolved problem in the heavy-ion collision.
We might as well say that chiral topological effect is the novel and
hottest unresolved challenge in the heavy-ion collision.  A marriage
of these investigations would produce fruitful offsprings.

%%%%% ACKNOWLEDGMENT %%%%%%%%

\vspace{1cm}

K.~F.\ thanks J\"{u}rgen~Berges, Jean-Paul~Blaizot, Francois~Gelis,
Alexi~Kurkela, Jinfeng~Liao, Larry~McLerran, Jan~Pawlowski,
S\"{o}ren~Schlichting, Mike~Strickland, Raju~Venugopalan for extremely
useful conversations, through which he learnt a lot.
This work was supported by Japanese MEXT grant (No.\ 15H03652 and
15K13479).

\section*{References}
\bibliographystyle{utphys}
\bibliography{evolve}

\end{document}